\documentclass[fleqn,usenatbib]{mnras}

\usepackage{newtxtext,newtxmath}

\usepackage[T1]{fontenc}

\DeclareRobustCommand{\VAN}[3]{#2}
\let\VANthebibliography\thebibliography
\def\thebibliography{\DeclareRobustCommand{\VAN}[3]{##3}\VANthebibliography}

\usepackage{graphicx}	
\usepackage{amsmath}	

\title[Spectroscopic analysis of BPS\,CS\,22940$-$0009]{Spectroscopic analysis of BPS\,CS\,22940$-$0009: connecting evolved helium stars}

\author[E.J. Snowdon et al.]{
E. J. Snowdon,$^{1}$\thanks{E-mail: edwardsnowdon@armagh.ac.uk}
L. J. A. Scott,$^{1}$
C. S. Jeffery,$^{1}$ and
V. M. Woolf,$^{2}$
\\
$^{1}$Armagh Observatory and Planetarium, College Hill, Armagh, BT61 9DB, UK\\
$^{2}$Physics Department, University of Nebraska at Omaha, 6001 Dodge St, Omaha, NE, 68182, USA
}

\date{Accepted XXX. Received YYY; in original form ZZZ}

\pubyear{2015}

\begin{document}
\label{firstpage}
\pagerange{\pageref{firstpage}--\pageref{lastpage}}
\maketitle

\begin{abstract}
BPS\,CS\,22940$-$0009 is a helium-rich B-star that shares characteristics with both helium-rich B subdwarfs and extreme helium stars. The optical spectrum of BPS\,CS\,22940$-$0009 has been analysed from SALT observations. The atmospheric parameters were found to be $T_{\rm eff} = 34\,970 \pm 370$ K, $\log g/{\rm cm\,s^{-2}} = 4.79 \pm 0.17$, $n_{\rm H}/n_{\rm He} \simeq 0.007$, $n_{\rm C}/n_{\rm He} \simeq 0.007$, $n_{\rm N}/n_{\rm He} \simeq 0.002$, although further improvement to the helium line fits would be desirable.  This places the star as a link between the He-sdB and EHe populations in $g$-$T$ space. The abundance profile shows enrichment of N from CNO-processing, and C from $3\alpha$ burning. Depletion of Al, Si, S and a low upper limit for Fe show the star to be intrinsically metal-poor. The results are consistent with BPS\,CS\,22940$-$0009 having formed from the merger of two helium white dwarfs and currently evolving toward the helium main sequence.
\end{abstract}

\begin{keywords}
stars: abundances -- stars: chemically peculiar -- stars: early-type -- stars: individual: BPS\,CS\,22940$-$0009 -- subdwarfs
\end{keywords}



\section{Introduction}

Hot subdwarf B (sdB) stars occupy the blue end of the horizontal branch and can be found in both the field and in globular clusters. They have colours typical of B stars and  effective temperatures and surface gravities similar to helium (He) main sequence stars of about 0.5 solar masses. Like other horizontal branch stars, they have helium-burning cores, but their hydrogen (H) envelopes are much thinner and cannot sustain shell burning. The spectra of sdBs are typically helium-deficient and show strong Balmer lines. It is thought that gravitational settling and radiative levitation sink the atmospheric helium below the hydrogen surface \citep{heber86}. However, about 4\% of sdBs display very helium-rich atmospheres characterised by strong neutral He lines \citep{green86}, with similar temperatures but lower surface gravities than their H-rich counterparts. Many of these helium-rich sdB stars (He-sdBs) also have strong lines of nitrogen (N {\sc ii} and N {\sc iii}) and in some cases, carbon (C {\sc ii} and C {\sc iii}) \citep{naslim10}. 

Extreme helium (EHe) stars are a rare class of peculiar supergiants with effective temperatures of 8\,000-32\,000\,K \citep{drilling84}. EHe spectra are characterised by strong lines of neutral He and singly-ionised C, with the Balmer lines being very weak or absent \citep{hunger75}. An overabundance of nitrogen in most EHe stars and a very high carbon abundance in all implies the presence of both H-processed and He-processed material at the surface \citep{heber83, jeffery96}. Two principal evolutionary models emerged. In the \textquoteleft double degenerate\textquoteright\,model \citep{webbink84,iben84,saio02}, a He white dwarf merges with a more massive C-O white dwarf companion. If the mass of the C-O component remains below the Chandrasekhar limit, the base of the accreted envelope ignites causing the envelope to expand. As the resulting H-deficient supergiant contracts, it will cool to become an EHe star and eventually a white dwarf. In the \textquoteleft final flash\textquoteright\,model \citep{iben83} a late ignition in the helium shell of a post-AGB star causes the outer layers to rapidly expand. Hydrogen in the envelope is consumed by burning and the resulting H-deficient supergiant contracts to become an EHe star. 

He-sdBs pose several questions in regards to their origins, evolution, and connection to similar populations. For example, how are they formed and what causes them to be so He-rich? Why are some stars abundant in C and/or N but not others? Do He-sdBs form a sequence with normal sdBs or with the He-rich O subdwarfs? Are there any links with other types of helium-rich stars such as R Coronae Borealis variables or helium white dwarfs \citep{jeffery08b}? To answer these questions, observing campaigns by (for example) \citet{naslim10, jeffery20} have focused on identifying helium-rich objects from surveys of faint blue stars, and systematically obtaining atmospheric parameters and chemical abundances from their spectra. This builds on earlier efforts to determine the physical parameters of He-sdBs using optical spectroscopy \citep[e.g.][]{ahmad03, ahmad04}. By collecting this information, it becomes possible to search for patterns within the He-sdB class and to identify possible  connections with other classes of evolved stars. 

Several scenarios have been postulated to explain the evolution of helium subdwarfs. \citet{brown01} proposed that a star undergoing a late helium-core flash while descending the white dwarf cooling curve could produce enhanced He and C abundances via flash mixing. \citet{lanz04} showed how flash mixing can be \textquoteleft deep\textquoteright \,or \textquoteleft shallow\textquoteright, depending on how deeply the hydrogen envelope is mixed into the site of the flash. Both kinds produce He-rich surfaces with enhanced C and N, but shallow mixing also leaves a significant remaining hydrogen fraction. These scenarios were expanded upon by \citet{millerbertolami08} who used 1D numerical simulations to study the effects of non-standard assumptions such as chemical gradients on hot-flasher mixing. Other authors have suggested that the coalescence of two helium white dwarfs (mergers) can account for both conventional and helium-rich subdwarfs. \cite{saio00} attempted to model such a merger by computing the consequences of rapid accretion of He-rich matter onto a helium white dwarf; this accretion results in off-centre helium ignition and causes the star to first expand, then contract through repeating helium-shell flashes. The resulting stellar surface is abundant in He and N, but shows no C enrichment. \citet{zhang12} simulated similar double helium white dwarf merger models assuming three combinations of rapid and slow accretion. They found that slow accretion allows the N-rich material of the accreted white dwarf to form the surface of the product without mixing. Fast accretion leads to convection zones that dredge up C-rich material to the surface. A combination of rapid (dynamical) followed by slow (thermal) accretion yields final abundances that depend on the mass of the merger, with high-mass systems producing deeper convective dredge-up after the first helium shell flash. Thus low-mass systems are expected to be N-rich, whilst high-mass systems are both N-rich and C-rich.  

BPS\,CS\,22940$-$0009 (EC 20262$-$6000) is a hot, He-rich star \citep{beers92} that is part of a distinct sequence of rare, H-poor stars with strong N lines. Similar objects include LS\,IV+6\textdegree2 \citep{jeffery98} and PG\,1415+492 \citep{drilling13}. These stars are typified by luminosity classes of V or less, and surface gravities similar to those of main sequence B stars. BPS\,CS\,22940$-$0009 was studied together with other N-rich He-sdBs by \cite{naslim10} and found to be the most C-rich member of the sample. Our goal for this work is to re-examine BPS\,CS\,22940$-$0009 using more recent spectra with improved signal-noise ratios in order to assess its relation to other He-sdB and EHe stars, and if it represents a connection between these classes. Additionally, we also aim to investigate the chemical profile of this star to search for any peculiarities. These could provide further clues to the evolution of this star and others like it.

In this paper, we present a spectroscopic analysis of BPS\,CS\,22940$-$0009. Section 2 describes the observations used for this paper. In section 3 we obtain the atmospheric parameters by fitting optical spectra to a grid of model atmospheres in local thermodynamic equilibrium (LTE), and calculate chemical abundances by measuring the equivalent widths of key spectral lines. We also use {\it TESS} photometry and cross-correlation of the HRS spectra to search for variability in flux and radial velocity. Section 4 discusses our abundance and parameter results and compares them to similar He-sdB and EHe stars.  The implications of our results on the evolutionary status of BPS\,CS\,22940$-$0009 are given in section 5. The potential impact of the LTE approximation is considered in Section \ref{sec:nlte}.

\section{Observations}

\begin{table}
    \centering
    \begin{tabular}{c|c|c|c|c|c}
    \hline
    Instrument/ & $R$/ & $t_{\rm exp}$ & S/N & Sampling\\
    Date & UT start & (s) & & (\AA/pix.) & \\
    \hline
    AAT/UCLES  & 32\,000 &    &     \\
    2005 08 26 & 09:25:40 & 1800 & 10 & 0.06\\
               & 09:56:33 & 1800 & 10 & 0.06\\
               & 10:27:29 & 1800 & 12 & 0.06\\
               & 10:58:22 & 1800 & 10 & 0.06\\
    ESO/FEROS  & 48\,000 &      &    &     \\
    2011 05 08 & 06:37:42 & 1800 & 20 & 0.6 \\
               & 07:08:35 & 1800 & 20 & 0.6 \\
    SALT/RSS   & 2250 &   &    &     \\
    2018 05 05 & 01:59:14 & 100  & 73 & 1.1 \\
               & 02:01:15 & 100  & 73 & 1.1 \\
               & 02:08:25 & 150  & 73 & 1.1 \\
               & 02:11:16 & 150  & 73 & 1.1 \\
    SALT/HRS   & 43\,000 &    &    &\\
    2016 06 23 & 02:49:43 & 1050 & 19 & 0.03\\
               & 03:08:24 & 1050 & 19 & 0.03\\
    2016 06 30 & 02:35:32 & 1250 & 20 & 0.03\\
               & 02:57:33 & 1250 & 20 & 0.03\\
    2018 05 07 & 01:00:12 & 900  & 16 & 0.03\\
               & 01:16:26 & 900  & 16 & 0.03\\
    2019 04 25 & 01:42:05 & 1450 & 38 & 0.03\\
               & 02:07:25 & 1450 & 38 & 0.03\\
    \hline
    \end{tabular}
    \caption{Record of observations of BPS\,CS\,22940$-$0009. The sampling of the reduced spectrum is given as an average in \AA/pixel over the full spectrum.}
    \label{tab:obslog}
\end{table}

BPS\,CS\,22940$-$0009 was selected for analysis due to its unique position as a bridge between stellar classes. This star was previously studied by \cite{naslim10}, who found it to be the coolest and lowest-gravity member of their sample. Additionally, a coarse analysis by \cite{jeffery20} placed the star as the hottest and highest-gravity member of a group of helium-rich stars that are not subdwarfs. 
With $\log g/{\rm cm\,s^{-2}} \approx 5$, it lies below the hydrogen main sequence ($\log g/{\rm cm\,s^{-2}} \approx 4$ for B stars).
This makes BPS\,CS\,22940$-$0009 a potentially interesting edge case among the He-sdB population, bridging the gap between more conventional hot subdwarfs and hot extreme helium stars. In recent years, further observations of BPS\,CS\,22940$-$0009 have made it possible to improve the accuracy of parameter and abundance measurements by increasing both resolution and signal-noise ratio.

Observations of BPS\,CS\,22940$-$0009 were made at the Southern African Large Telescope ({\it SALT}). Detailed optical spectra were obtained using the High Resolution Spectrograph ({\it HRS}: $R\simeq43\,000,\,\lambda\lambda=4100-5200$\,\AA) which is an \'echelle format spectrograph. Since broad-lines frequently span more than one \'echelle order, targets are also observed with the lower-resolution Robert Stobie Spectrograph ({\it RSS}: $R\simeq 2000,\,\lambda\lambda=3850-5150$\,\AA) in order to guide the correction of the \'echelle blaze function.

The HRS spectra for this analysis were obtained on the night of 24-25 April 2019 and consisted of 4x1450 second exposures; 2 red and 2 blue. The RSS observations were performed on the night of 4-5 May 2018 and consisted of 2x100 second and 2x150 second exposures. We used the PG2300 grating and a slit width 1.5\AA\, to obtain a full-width half-maximum resolution $R$ between 1800 at 4000\AA\, and 2300 at 5000\AA\ (as measured from the CuAr calibration lamp),  corresponding to $\Delta \lambda \approx 2.2$\AA\,  throughout. The RSS detector subsystem comprises 3 CCD chips separated by 2 gaps, so double exposures are taken from 2 different grating angles and merged to produce a continuous spectrum. The data reduction process for the RSS spectra is described by \citet{jeffery20}. The HRS data were reduced using standard {\sc iraf} routines to subtract the bias signal, divide stellar spectra by combined flat field spectra, correct pixels affected by cosmic rays, reduce to one dimensional spectra, and apply the wavelength scale using thorium-argon lamp spectra. Spectra from previous observations were also used to search for variability in radial velocity over time. The full list of available observations is given in Table \ref{tab:obslog}. A combined spectrum was produced from a S/N-weighted average of all available HRS observations. This offered the highest-available signal-noise ratio (S/N = 52), but did not produce a significantly better fit to interpolated grid models than the 2019 spectrum alone ($\chi^2_{\rm all}/\chi^2_{2019} = 0.89$).
Furthermore, including the 2016 and 2018 observations introduced noise in the échelle overlap regions of the combined spectrum. As a result, only the 2019 spectrum was used for the majority of the analysis.

\section{Atmospheric Analysis}

The atmospheric parameters effective temperature ($T_{\rm{eff}}$), surface gravity ($g$), and fractional helium abundance by number ($n_{\rm{He}}$) were determined simultaneously using the software package {\sc sfit} \citep{jeffery01}, which finds the best match to the observed spectrum in a grid of theoretical spectra. Chemical abundances were calculated from the equivalent widths of absorption lines in the HRS spectrum. 

\subsection{Model atmospheres}
Grids of model atmosphere structures were constructed using the model atmosphere code {\sc sterne} \citep{behara06}. The models assume local thermodynamic equilibrium, hydrostatic equilibrium, and use plane-parallel geometry. Grids of high-resolution synthetic spectra were computed therefrom using the LTE formal solution code {\sc spectrum} \citep{jeffery01}. Subgrids were created by resampling the master grid onto a wavelength interval of 1.1\,\AA. This grid would be used for fitting the RSS spectrum. Both grids covered a parameter space of $T_{\rm eff}$/K = 28\,000 (2\,000) 40\,000, log $g$/cm\,s$^{-2}$ = 4.0 (0.25) 5.5 and $n_{\rm He} = 0.95, 0.99, 1.0$. 

\begin{figure*}
    \centering
    \includegraphics[width=15cm]{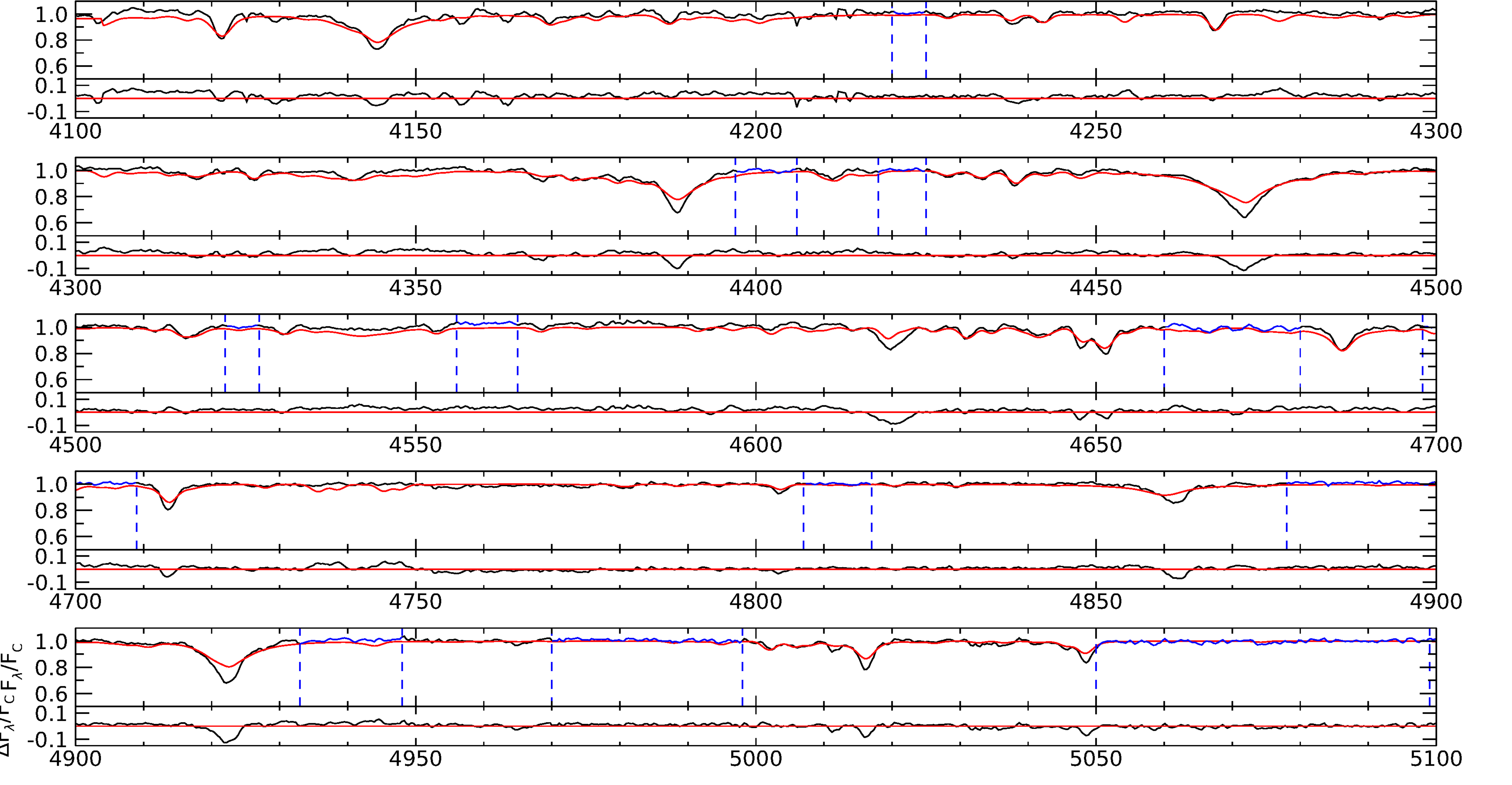}
    \caption{Comparison of the renormalised RSS spectrum (black) and a model spectrum (red), which has $T_{\rm eff}$ = 34\,970 K, log $g$ = 4.79, $n_{\rm He}$ = 0.978. Designated continuum regions used for renormalisation in the fitting process are shown in blue. The residual flux (observed -- model) is shown beneath.}
    \label{fig:fit}
\end{figure*}

\subsection{Atmospheric parameters}

{\sc sfit} interpolates in a grid of model spectra using a set of initial guesses for the atmospheric parameters $T_{\rm{eff}}$, $g$ and $n_{\rm{He}}$. The interpolated spectrum is compared with the observed spectrum by evaluating the $\chi^2$ statistic. A best-fit solution is found using a Levenburg-Marquardt algorithm to iterate the guess parameters until $\chi^2$ is minimised \citep{press89}.
The atmospheric parameters were first measured for the RSS spectrum. Reference regions relatively free of absorption lines and distributed across the full wavelength range (4100-5100\,\AA) were defined manually. 
A reference spectrum was defined using initial parameter estimates $T_{\rm eff}=34\,000$\,K, $\log g/{\rm cm\,s^{-2}} = 4.5$, $n_{\rm He}=0.99$. The reference regions were used with a high-pass filter having FWHM 200\,\AA\ to renormalise the observed spectrum to the reference spectrum. {\sc sfit} then solved for $T_{\rm eff}$, $\log g$, and $n_{\rm He}$ individually, assuming an instrumental broadening of 1\,\AA. $T_{\rm eff}$ was measured by fitting the temperature-sensitive He\,{\sc ii} 4686\,\AA\,line. $\log g$ was measured from the gravity-sensitive He\,{\sc i} 4388, 4471, and 4921\,\AA\,lines. The cores of these lines were excluded from the fit as they are not well-reproduced by the LTE model. $n_{\rm He}$ was obtained by fitting the whole spectrum; {\sc sfit} calculates the hydrogen abundance of the fit, and then reports $n_{\rm He} = 1 - n_{\rm H}$.

The parameter results were found to be $T_{\rm eff}=34\,970 \pm 370$\,K, $\log g/{\rm cm\,s^{-2}}=4.79 \pm 0.17$, $n_{\rm He}=0.978\pm0.006$ (by number) from the RSS spectrum and $T_{\rm eff}=34\,420 \pm {240}$\,K, $\log g/{\rm cm\,s^{-2}}=4.66 \pm 0.10$, $n_{\rm He}=0.974\pm0.006$ from HRS (Table\,\ref{tab:params}). Errors are the formal {\sc sfit} errors obtained by solving for each parameters individually, with the others held constant. The small error on $n_{\rm He}$ arises because this is essentially a measurement of $1-n_{\rm H}$ which is tightly constrained by the Balmer lines. A comparison of the normalised RSS spectrum and a model spectrum created with these parameters is shown in Fig.\,\ref{fig:fit}. The model atmosphere having parameters closest to those of the best-fitting RSS spectrum was then used to measure $v_{\rm t}$, and $v{\rm sin}\,i$ from the HRS spectrum (\S\,3.3), and the stellar radius, mass, and luminosity from the {\it TESS} spectral energy distribution (\S\,3.7).

\begin{table*}
    \centering
    \begin{tabular}{c|c|c|c|c}
        \hline
         & SALT/RSS & SALT/HRS & Naslim 2010 & Jeffery 2020\\
        \hline
        $T_{\rm eff}$ (K) & $34\,970 \pm 370$ & $34\,420 \pm 240$ & 33\,700 $\pm$ 800 & 34\,650 $\pm$ 110\\
        log $g$ (cm\,s$^{-2}$) & $4.79 \pm 0.17$ & $4.66 \pm 0.10$ & 4.7 $\pm$ 0.2 & 4.89 $\pm$ 0.04\\
        $n_{\rm He}$ & 0.978 $\pm$ 0.006 & 0.974 $\pm$ 0.06 & 0.993 & 0.98 $\pm$ 0.01\\
        $v_{\rm rad}$ (km\,s$^{-1}$) & - & 32.7 $\pm$ 2.0 & -- & 28 $\pm$ 3\\
        $v_{\rm t}$ (km\,s$^{-1}$) & - & 7.8 $\pm$ 0.2 & 10 & --\\
        $v$sin$i$ (km\,s$^{-1}$) & - & 15.0 $\pm$ 3.3 & 4 $\pm$ 3 & --\\
        \hline
    \end{tabular}
    \caption{Atmospheric parameters for the spectrum of BPS\,CS\,22940$-$0009 obtained using {\sc sfit}. Results from \citep{naslim10} and \citep{jeffery20} are shown for comparison.}
    \label{tab:params}
\end{table*}

\subsection{Microturbulent velocity and chemical abundances}

Using the above parameters, the model atmosphere corresponding to the best-fit spectrum was selected from the grid. Using {\sc spectrum}, predicted equivalent widths ($W_{\lambda}$) were computed for all lines in the observed spectral window. Lines predicted to have $W_{\lambda}\geq5$\,m\AA\  were identified from this list; equivalent widths were then measured for all qualifying C, N, O, Ne, Mg, Al, Si, and S lines observable in the HRS spectrum. A complete list of all line measurements made for the abundance calculation is provided in Appendix \ref{app:linelist}.

For each line the part of the profile to be measured was defined manually (the segment), along with a region of continuum on either side of the line. A linear fit was computed from the continuum regions, and the equivalent width was obtained by integrating the line segment under the continuum fit. A parabola was fit to the line segment to  provide the  line wavelength. Errors were derived from an estimate of the flux error in the continuum regions and from the formal parabola fit.
 
\begin{figure}
	\includegraphics[width=\columnwidth]{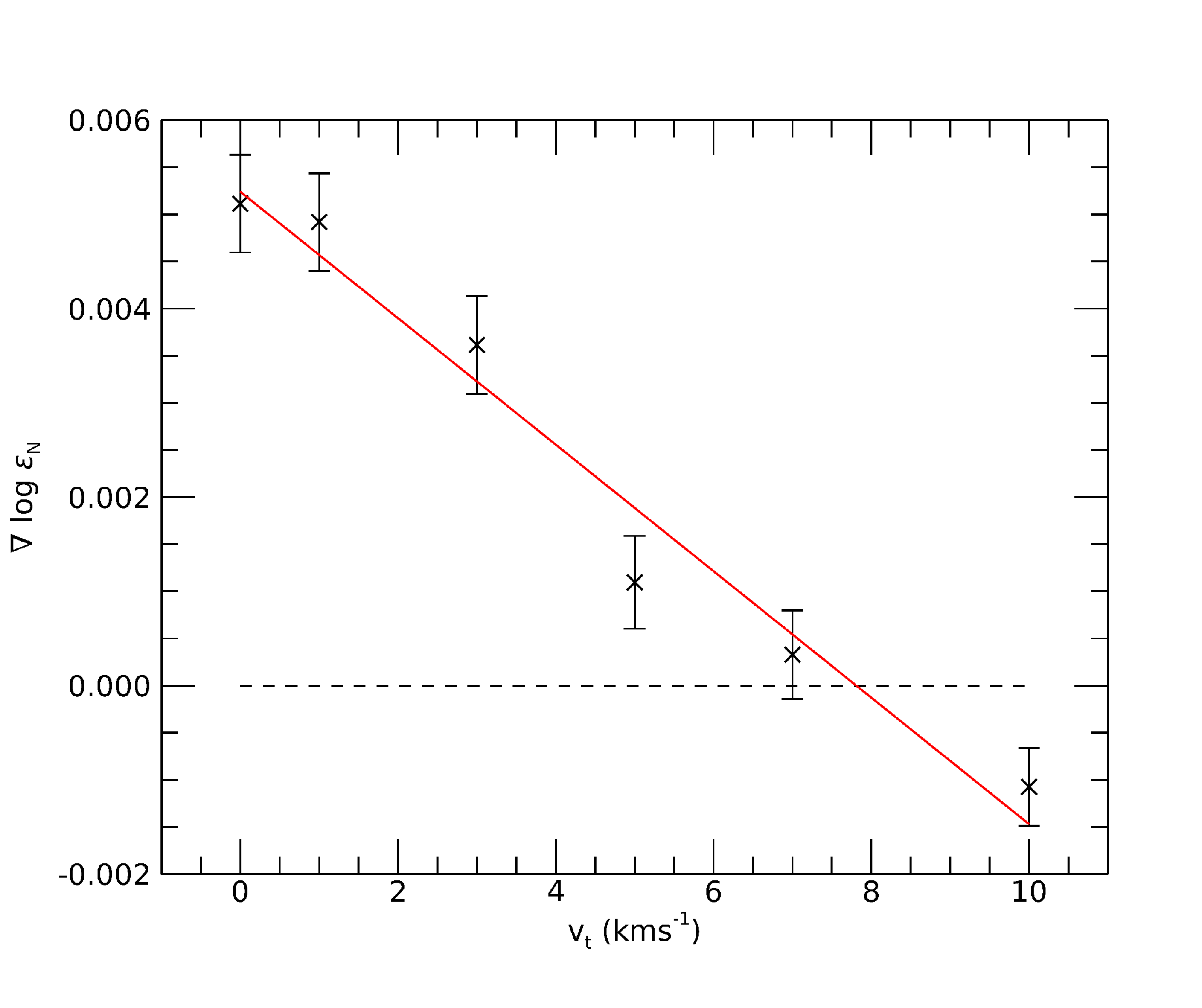}
    \caption{Gradient of nitrogen abundance as a function of equivalent width vs assumed microturbulent velocity for 36 N{\sc ii} lines. The red line is a linear fit that crosses $\nabla\log\epsilon_{\rm N}=0$ at $v_{\rm t}=7.8 \pm 0.2{\rm km\,s^{-1}}$ with $\chi^2=1.12x10^{-6}$.}
    \label{fig:vturb_regression}
\end{figure}

For a given (assumed) microturbulent velocity ($v_{\rm t}$), {\sc spectrum} may be used to compute a predicted line equivalent width for a given chemical abundance or, by Newton-Raphson iteration, the chemical abundance corresponding to a measured equivalent width. Nitrogen lines were used to determine $v_{\rm t}$ using the classical method of minimising the gradient of abundances with respect to equivalent width \citep{gray75}. Abundances were calculated for each of 36 N {\sc ii} lines with assumed $v_{\rm t}$ between 0 and 10 km\,s$^{-1}$. The abundance-equivalent width gradients ($\nabla\log\epsilon_{\rm N} = d\log\epsilon_{\rm N}/dW_{\lambda}$) were determined by $\chi^2$ minimisation. Only lines giving abundances of $8.0 \le \log\epsilon_{\rm N} \le 10.0$ were considered to avoid errors due to saturated, weak ($W_{\lambda}<5$\,m\AA) or blended lines. A linear fit of the gradients gives $\nabla\log\epsilon_{\rm N}=0$ for $v_{\rm t}=7.8 \pm 0.2\,{\rm km\,s^{-1}}$ (Fig.\, \ref{fig:vturb_regression},  Table\,\ref{tab:params}).
 
Adopting $v_{\rm t}=7.8$, chemical abundances were derived for all lines with measured $W_{\lambda}\geq5$\,m\AA. Rejecting outliers, mean abundances were derived for 8 species (C, N, O, Ne, Mg, Al, Si, and S: Table\,\ref{tab:ew_abunds}). Abundances are given in the form $\log \epsilon_{i} = \log n_{i} +c$, where $c=11.5725$ and $n_{i}$ are number fractions\footnote{Stellar abundances are conventionally given such that for hydrogen $\log\epsilon_{\rm H}=12$, and for other species $i$ by $\log\epsilon_{i} = \log n_i/n_{\rm H}$.} This assumes a hydrogen-dominated composition, which is not appropriate for helium-rich stars, like BPS\,CS\,22940$-$0009, where a) the denominator may approach zero and b) number fractions of conserved species change simply because 4 protons combine to form a single helium nucleus. With atomic weights $a_{i}$, $c$ is computed such that $\log \Sigma_{i} a_{i} n_{i} + c = \log \Sigma_{i} a_{i} n_{i,\odot} + c_{\odot} = 12.15$. This conserves both $\log \epsilon_{i}$ and mass fraction for each species whose abundance is otherwise unchanged \citep{jeffery11}.. The error in the mean abundances was propagated quadratically from the errors in the individual abundances in each line, and hence from the error in the equivalent width measurements. Upper limits were obtained for P and Fe. Individual species are discussed below.

\begin{table}
	\centering
	\caption{Chemical abundances for BPS\,CS\,22940$-$0009. Abundances are given in the form $\log \epsilon_{i} = \log n_{i} +c$ as defined in the text. $N$ is the number of lines used in the abundance calculation. Abundance errors for metals are propagated from the errors in the individual line measurements reported by {\sc spectrum}. Errors for H and He are taken from the fractional error in $n_{\rm He}$ and $n_{\rm H}$. Results from \citet{naslim10} and photospheric solar values from \citet{asplund09} are included for comparison. These are normalised such that $\log \Sigma_{i} a_{i} n_{i}=12.15$.}
	\label{tab:ew_abunds}
	\begin{tabular}{lcccc} 
		\hline
		Species & $N$ & $\log\epsilon$ & $\log\epsilon$ (Naslim) & $\log\epsilon$ (Solar) \\
		\hline
		H & - & 9.91 $\pm$ 0.06 & 9.10 $\pm$ 0.20 & 12.00\\
		He & - & 11.56 $\pm$ 0.01 & 11.54 & 10.93 $\pm$ 0.01\\
		C & 45 & 9.43 $\pm$ 0.68 & 8.94 $\pm$ 0.35 & 8.43 $\pm$ 0.05\\
		N & 69 & 8.85 $\pm$ 0.66 & 8.46 $\pm$ 0.22 & 7.83 $\pm$ 0.05\\
		O & 9 & 7.31 $\pm$ 0.45 & 7.11 $\pm$ 0.34 & 8.69 $\pm$ 0.05\\
		Ne & 15 & 7.93 $\pm$ 0.47 & 8.27 $\pm$ 0.45 & 7.93 $\pm$ 0.10\\
		Mg & 3 & 7.81 $\pm$ 0.41 & 7.27 $\pm$ 0.18 & 7.60 $\pm$ 0.04\\
		Al & 6 & 6.01 $\pm$ 0.67 & 6.12 $\pm$ 0.15 & 6.45 $\pm$ 0.03\\
		Si & 5 & 6.98 $\pm$ 0.28 & 7.23 $\pm$ 0.24 & 7.51 $\pm$ 0.03\\
		P & - & $\leq$ 5.25 & - & 5.41 $\pm$ 0.03\\ 
		S & 4 & 6.25 $\pm$ 0.43 & 6.45 $\pm$ 0.15 & 7.12 $\pm$ 0.03\\
		Fe & - & $\leq$ 6.30 & - & 7.50 $\pm$ 0.04\\
		\hline
	\end{tabular}
\end{table}

The abundances obtained by the equivalent width method were used in {\sc spectrum} to construct a formal solution for the whole spectrum. A comparison of the synthetic spectrum and the re-normalised HRS spectrum is shown in Appendix \ref{app:hrs_spec}.

\subsubsection{H and He}

The spectrum of BPS\,CS\,22940$-$0009 is dominated by very strong helium lines, while the Balmer lines are comparatively weak and blended with weak He\,{\sc ii}. The LTE models do not fit the cores of these lines well; the LTE assumption is discussed in Section \ref{sec:nlte}, in which a comparable NLTE model shows overpopulation of the lower levels of the He\,{\sc i} line transitions, leading to deeper line cores than the LTE case.
The gravity-sensitive He\,{\sc i} (and He\,{\sc ii}) lines have widths comparable with those of individual orders in the HRS \'echelle. Moreover, blaze removal and order merging in the HRS data reduction process remain imperfect. Consequently, the wings of broad lines in the HRS spectrum may be unsuitable for measuring $T_{\rm eff}$, $\log g$ and $n_{\rm He}$. The final He and H abundances were therefore obtained via $\chi^2$ minimisation using the RSS spectrum. Fractional abundances $n_{\rm H} = 0.022 \pm 0.003$ and $n_{\rm He} = 0.978 \pm 0.006$ yield $\log\epsilon_{\rm H} = 9.40 \pm 0.06$ and $\log\epsilon_{\rm He} = 11.56 \pm 0.01$ as defined above. The H abundance was measured from the H$_{\beta}$ and H$_{\gamma}$ lines as a check on the results from {\sc sfit}, giving $\log\epsilon_{\rm H} = 9.44 \pm 0.07$. The atmosphere is strongly He-rich and H-poor, with $n_{\rm H}/n_{\rm He}\simeq0.007$. This is typical for He-sdB and EHe stars, though hydrogen is less depleted than in the similar stars LS\,IV+6\textdegree2 \citep{jeffery98} and BX\,Cir \citep{drilling98}.

\subsubsection{C, N, and O}

The spectrum contains a large number of carbon and nitrogen lines with a wide range of strengths, including several that are blended with other lines and many that are saturated. Taking the abundance as a simple mean of all line measurements is not viable as blended lines skew the average abundance to a higher apparent value. Instead, the modal abundance from all lines with equivalent widths $\geq5$\,m\AA\, was taken as a starting point. All lines with a measured abundance outside of one standard deviation from the mode were excluded. The overall abundance was then taken to be the mean of the remaining lines, with the error being propagated forward from the errors in each abundance measurement reported by {\sc spectrum}. For carbon, this resulted in an abundance of $\log\epsilon_{\sc \rm C} = 9.43 \pm 0.68$, based on measurements of 45 lines. For nitrogen, the result was $\log\epsilon_{\sc \rm N} = 8.85 \pm 0.66$ based on 69 lines.

There is a broad feature overlapping the C\,{\sc II} lines at $\sim$4618\,{\AA}. These lines arise from the transition between electronic states which can both autoionise to the ground state of C\,{\sc III}. This discrete-to-continuum transition produces the characteristic broad feature whose absorption cross section can be fitted by a Fano profile \citep{fano1961}, as in e.g. \citet{yan1987}. There was no need to attempt to fit this feature in our case since the other carbon lines were adequate for abundance measurement. There are also a number of C\,{\sc II} lines in the 4720 to 4760\,{\AA} region which appear in the modelled LTE spectrum but are absent in the data. These lines are known to be sensitive to NLTE effects and are partially in emission.
Both features are well known in the spectra of carbon-rich and hydrogen-deficient stars of appropriate temperature and surface gravity
\citep{klemola61, heber86.hdef1a, leuenhagen94a}.
  
The spectrum lacks strong oxygen lines. The oxygen abundance was measured from 9 relatively distinct lines. These gave a mean abundance of $\log \epsilon_{\sc \rm O} = 7.31 \pm 0.45$.

\subsubsection{Other metals}

The neon abundance was measured from 15 lines between 4219.74-4522.72\,\AA, giving a mean result of $\log \epsilon_{\sc \rm Ne} = 7.93 \pm 0.47$. The only magnesium feature in the spectrum is a blend of 3 lines at 4481\,\AA. The Mg abundance from this blend had to be measured carefully, since it lies within the wing of a broad He line which causes curvature in the pseudo-continuum. The equivalent width measurement was repeated 3 times, keeping the designated continuum as close to the line as possible. This gave a mean Mg abundance of $\log \epsilon_{\sc \rm Mg} = 7.81 \pm 0.41$. Using this value in the model reproduced the width of the line well, but the core was too shallow. However, the Mg line is saturated, so increasing the abundance in the model caused the wings to broaden and reduced the accuracy of the fit. This suggests a supersolar Mg abundance, but without any other Mg features to measure this cannot be determined conclusively.

The aluminium abundance was measured from 6 lines between 4149.90 and 4529.20\,\AA\, to be $\log \epsilon_{\sc \rm Al} = 6.01 \pm 0.67$. Similarly, the sulphur abundance was found using lines at 4253.59, 4284.98, 4332.69, and 4361.53\,\AA\, to be $\log \epsilon_{\sc \rm S} = 6.25 \pm 0.43$.

The spectrum contains both strong and weak silicon lines. The shapes of the strong lines are not well-reproduced by the LTE model, so the abundance was measured from weaker lines, specifically 4212.41\,\AA\, and the four lines between 4716-4830\,\AA. The mean Si abundance was found to be $\log \epsilon_{\sc \rm Si} = 6.98 \pm 0.28$.

The iron and phosphorus abundances are important for determining the overall metallicity, but neither species has lines with $W_{\lambda}\geq5$\,m\AA\, within the 4100-5100\,\AA\, window. An upper limit on these abundances was established by finding the minimum abundances at which the strongest lines in the window (Fe {\sc iii} 4164.73\,\AA\, and P {\sc iii} 4222.20\,\AA) become detectable to a confidence level of 68\% (Appendix \ref{app:width}). These were found to be $\log \epsilon_{\sc \rm Fe} \leq 6.30$ for iron and $\log \epsilon_{\sc \rm P} \leq 5.35$ for phosphorus. The S/N-weighted average spectrum from all available HRS observations was used to measure the Fe upper limit as it provided an improved signal-noise ratio in the region around Fe {\sc iii} 4164.73\,\AA. P {\sc iii} 4222.20\,\AA\, lay in a noisy échelle overlap region and so the P upper limit was measured using only the 2019 HRS spectrum.

Taking the Al, Si, and S abundances as a proxy of metallicity, BPS\,CS\,22940$-$0009 is metal poor by $\simeq 0.6 \pm 0.3$ dex on average compared to solar values \citep{asplund09}. These measurements are $\simeq 0.2 \pm 0.2$ dex lower on average than those from \cite{naslim10}, with the results being in agreement for most species. The noisier AAT/UCLES spectrum makes it very difficult to measure the equivalent widths of the weak lines, so the abundances were based on less sensitive saturated lines. The Mg blend is also strongly affected by noise in the continuum and in the blue wing of He 4471.48\,\AA. This makes the equivalent width of the feature harder to measure in the UCLES spectrum than in the HRS spectrum and may explain the difference in Mg abundance results. Similarly, the weak Si lines are not easily measurable in the UCLES spectrum, which may explain the significant differences in Si abundance.

\subsection{Rotational Broadening}

The projected equatorial velocity ($v \sin i$) was assumed to be 0\,km\,s$^{-1}$ for the RSS spectrum as this would not have a significant effect due to the low resolution. For the HRS spectrum, {\sc sfit} was used to measure $v \sin i$ by fitting a formal solution to the spectrum in the region 4770--4830\,\AA\,with all other parameters fixed to their values in the RSS solution. This region has a high signal-noise ratio and contains metal lines (3 Si {\sc iii}, 7 N {\sc ii}, 1 C {\sc ii}/N {\sc ii} blend) that are sensitive to $v \sin i$ broadening. This gave a result of $v \sin i = 15.0 \pm 3.3\,{\rm km\,s^{-1}}$.

\begin{figure}
    \centering
    \includegraphics[origin=c, width=8.5cm]{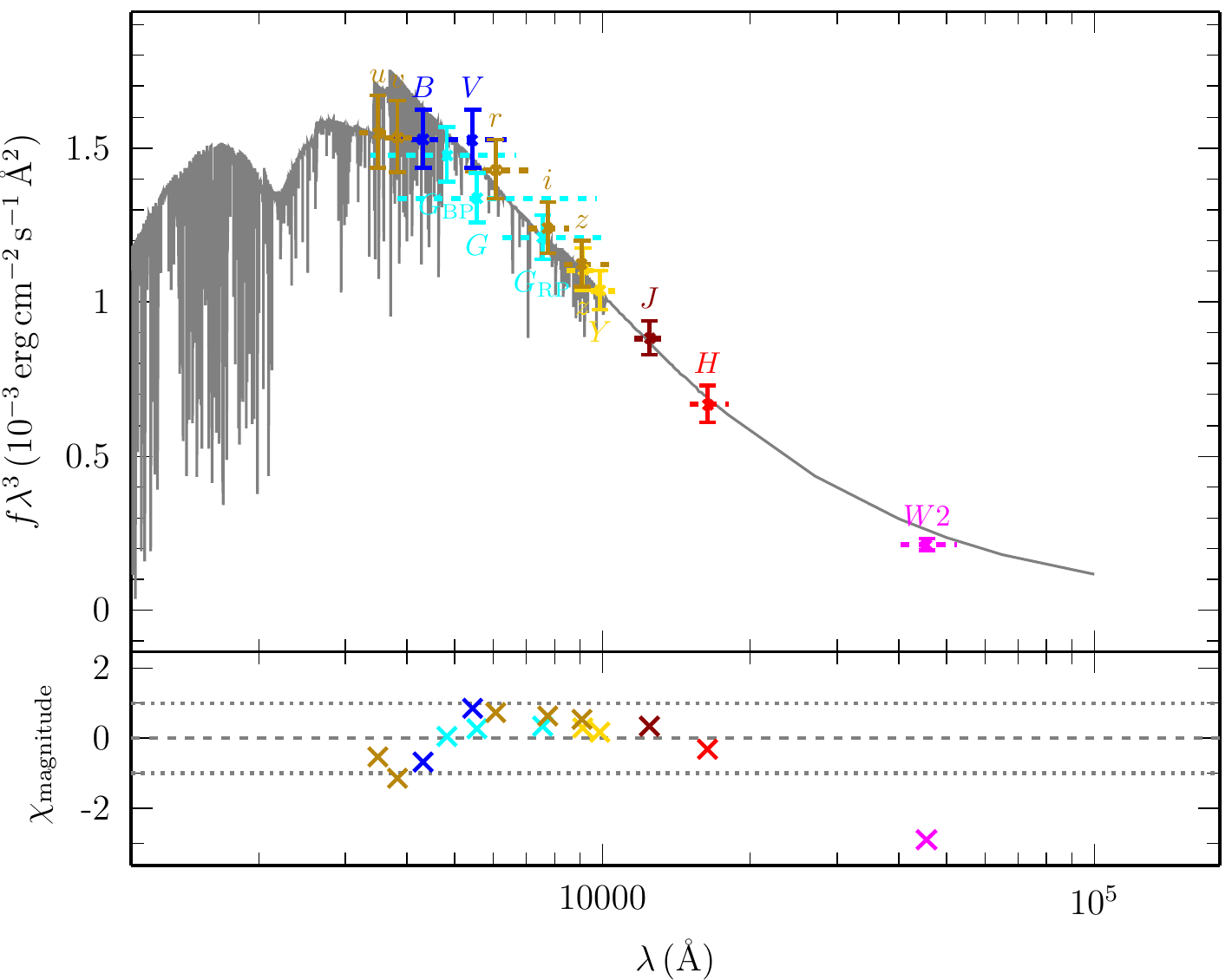}
    \caption{Flux distribution of the best-fit model atmosphere with $T_{\rm eff} = 34\,970 $K normalised to broadband photometry of BPS CS 22940$-$009. The lower panel shows the residuals normalized to the individual measurement error. Photometry is obtained from the following catalogues \citet{skymapper_dr2} ($u,v,r,i,z$), 
    \citet{apass9} ($B,V$), 
    \citet{gaiaedr3} ($G_{\rm BP},G,G_{\rm RP}$), 
    \citet{des_dr1} ($Y$),
    \citet{vhs_source} ($J$), 
    \citet{twomass} ($H$),  and
    \citet{unwise} ($W2$).  }
    \label{fig:phot}
\end{figure}

\subsection{Spectral Energy Distribution}

The stellar radius ($R$), mass ($M$) and luminosity ($L$) were obtained by fitting the reddened spectral energy distribution (SED) of the best-fit model atmosphere ($T_{\rm eff}$ / K = 34\,970, $\log g/{\rm cm\,s^{-2}} = 4.79$) to observed visual and near-infrared photometry (Fig.\,\ref{fig:phot}), using the {\sc isis} SED fitting toolkit \citep{heber18,irrgang21}. 
The reddening was calculated using the extinction law of \citet{fitzpatrick19} with $E(44-55) = 0.060 \pm 0.015$. 
The fit then yielded an angular diameter $\log \theta /{\rm rad} = -11.054 \pm 0.008.$ (where $\theta = 2R/d$). 
Using the {\it Gaia} EDR3 parallax of $0.47\pm 0.04$\,mas and subtracting a zero-point correction of $-26\,{\rm \upmu as}$ \citep{lindegren21}, we obtain $d = 2.10^{+0.19}_{-0.16}$\,kpc and thus $R = 0.42 \pm 0.04$\,${\rm R_\odot}$. Using $R$ with $T_{\rm eff} = 34\,970 \pm 370 $\,K from the RSS spectrum, we obtain $L = 230^{+50}_{-40}\,{\rm L_\odot}$. 
Using $R$ with $\log g / {\rm cm\,s^{-2}}=4.79 \pm 0.17$, we obtain $M = 0.39^{+0.08}_{-0.06}\,{\rm M_\odot}$. 
To indicate the impact of systematic errors, the SED fit obtained using the HRS parameters (Table\,\ref{tab:params}) yields $L = 210^{+50}_{-40}\,{\rm L_\odot}$ and $M = 0.29^{+0.06}_{-0.05}\,{\rm M_\odot}$.

\subsection{Radial velocity}

\begin{table}
    \centering
    \begin{tabular}{c|c|c}
    \hline
        Date & $v_{\rm rad}$ & $\delta v_{\rm rad}$\\
             & (km\,s$^{-1}$)  & (km\,s$^{-1}$)\\
        \hline
        2016 06 23 & 24.6 & 0.9\\
        2016 06 30 & 22.8 & 1.7\\
        2018 05 07 & 31.5 & 0.8\\
        2019 04 25 & 32.7 & 2.0\\
        \hline
    \end{tabular}
    \caption{Heliocentric radial velocities from cross-correlation of each available HRS observations of BPS\,CS\,22940$-$0009 with the best-fit model spectrum.}
    \label{tab:vrads}
\end{table}

The radial velocity ($v_{\rm rad}$) was determined by cross-correlation of the HRS spectrum and the best-fitting model spectrum. The peak of the cross-correlation function provided a wavelength shift that was used to calculate $v_{\rm rad}$ after applying a barycentric correction. The error in $v_{\rm rad}$ was calculated by measuring $\chi^2$ between the model spectrum and the observed spectrum across the range of wavelength shifts used for the cross-correlation function. The $1\sigma$ error bar was taken to be where $\Delta\chi^2=\chi^2-\chi^2_{\rm min}=1$.

Radial velocities were measured for all 4 available HRS observations of BPS\,CS\,22940$-$0009, to investigate any possible variability. Table\,\ref{tab:vrads} shows a distinct difference in  $v_{\rm rad}$ between the 2016 and 2018/2019 spectra. However, pre-2018 HRS observations have shown significant velocity errors associated with the HRS calibration programme \citep{jeffery19}. Therefore it is unclear whether there is real variability in the $v_{\rm rad}$ of BPS\,CS\,22940$-$0009.

\subsection{Photometry}

\begin{figure*}
\includegraphics[width=12cm]{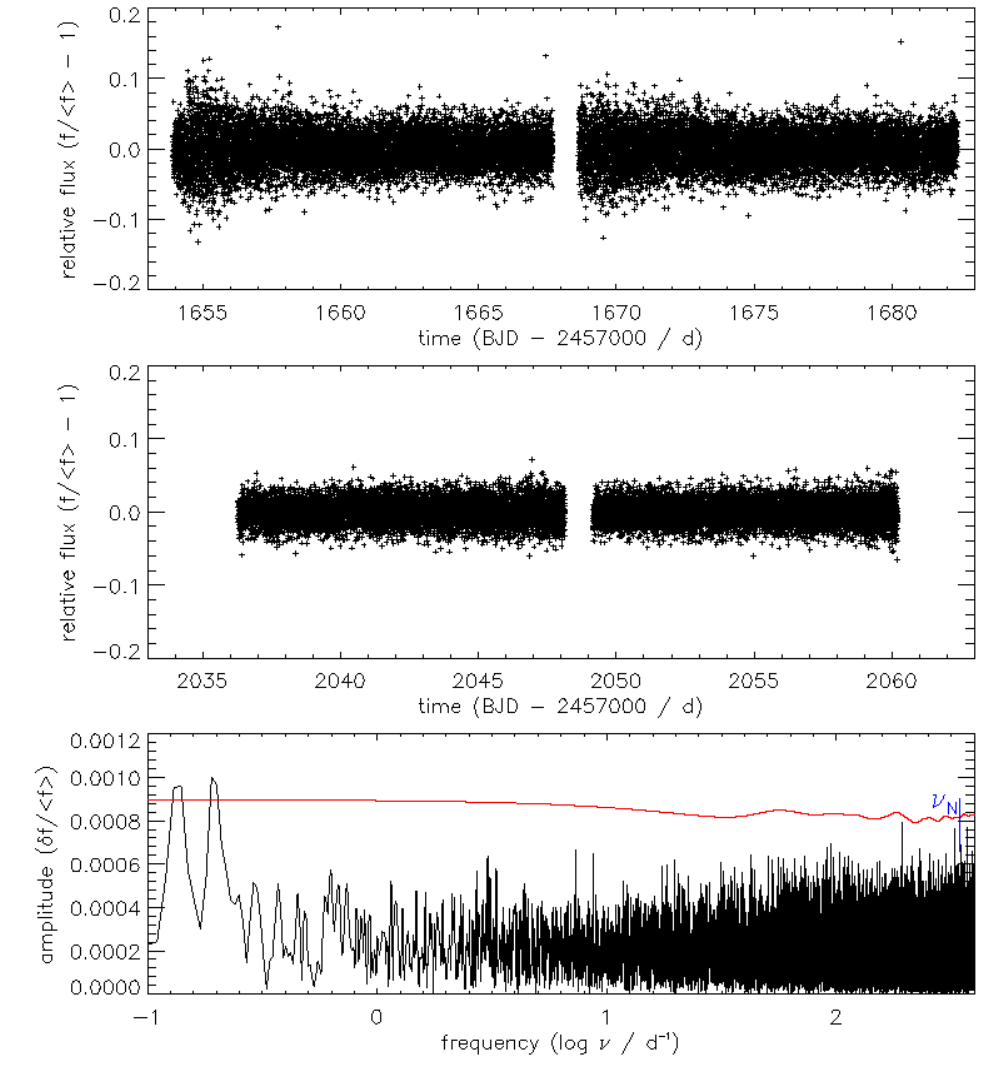}
\caption{{\it TESS} photometry of BPS\,CS\,22940$-$0009 during Sectors 13 (top) and 27 (middle) expressed as differential flux relative to the mean,
and as a frequency amplitude spectrum (bottom).
The Nyquist frequency ($\nu_{\rm N}$: 2\,m cadence) is marked by a blue vertical bar.
The red line shows the 4$\sigma$ level above which discrete frequencies might be considered real.}
\label{fig:tess}
\end{figure*}

BPS\,CS\,22940$-$0009 was observed in 2\,m cadence in a 600-1000\,nm bandpass with the Transiting Exoplanet Survey Satellite ({\it TESS}) during Sectors 13 and 27 (Fig.\,\ref{fig:tess}). In relative flux units, the mean individual datum error is 2.9\%. The standard deviation of the data is 2.2\% in a total of 17\,546 observations. The Fourier transform was investigated for periodic content (Fig.\,\ref{fig:tess}). For frequencies ($\nu$) greater than 0.25 d$^{-1}$ (periods less than 4 days) there is no evidence for periodic variability with (semi-) amplitude $a > 0.06\%$.
Low-frequency signal at $\nu<0.25\,{\rm d}^{-1}$ with $a\approx 0.1\%$ is likely to be red noise.

\section{Discussion}

\subsection{Atmospheric parameters}

\begin{figure*}
    \centering
    \includegraphics[width=\textwidth]{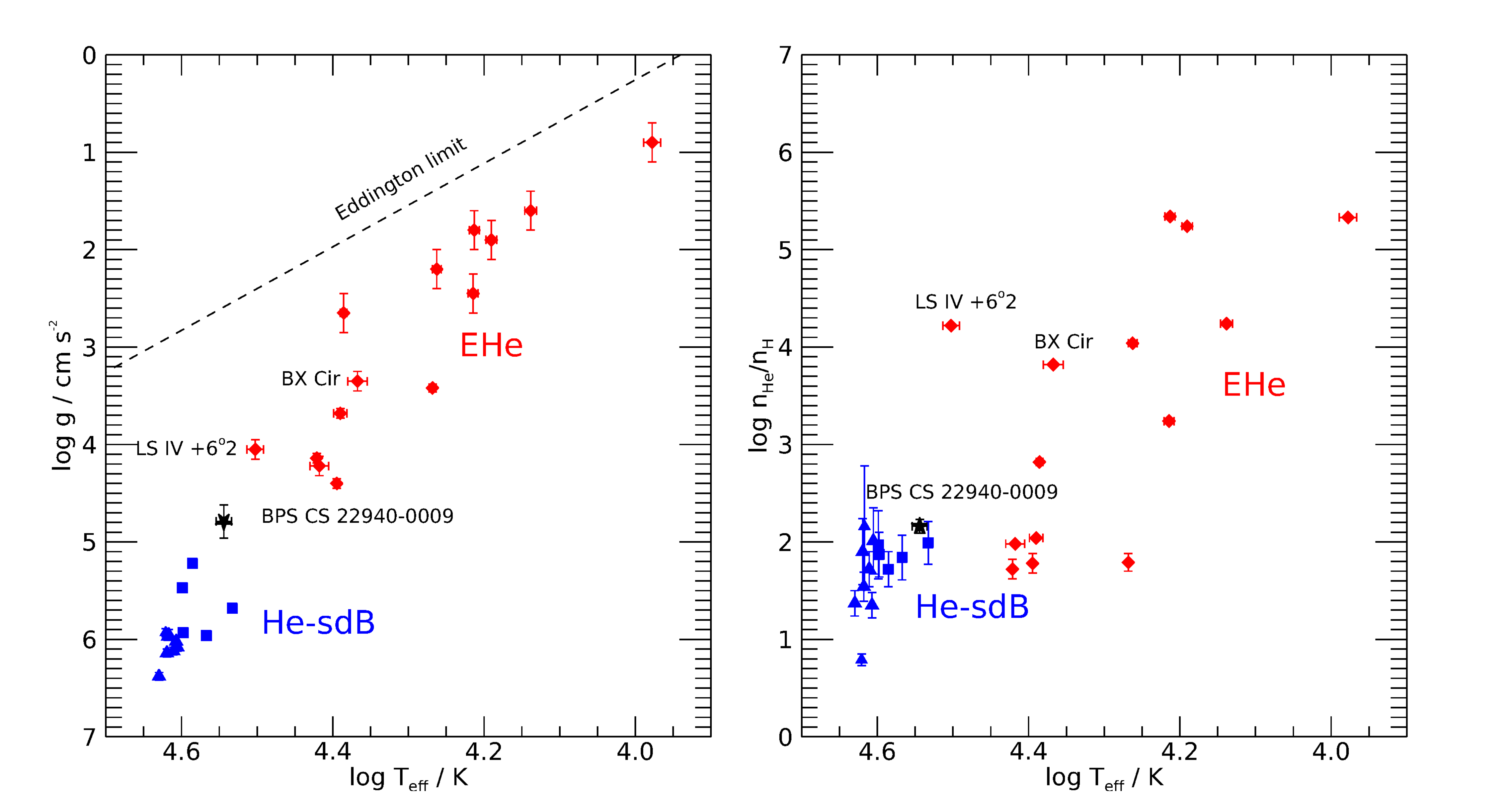}
    \caption{Comparison of the surface gravity (left) and He/H abundance ratio (right) against effective temperature of BPS\,CS\,22940$-$0009 (black) versus He-rich sdB (blue) and extreme helium (red) stars. Blue squares indicate He-sdBs with strong C lines in their spectra (He-sdBC). Blue triangles indicate He-sdBs with weak or absent carbon. The positions of the comparable stars LS\,IV\,+6\textdegree2 and BX\,Cir are indicated. Parameters for comparison stars taken from \citet{jeffery20,jeffery08b}. The Eddington limit for Thomson scattering for H-deficient atmospheres is included in the left panel \citep{mihalas78}.}
    \label{fig:param_comp}
\end{figure*}

The full list of atmospheric and stellar parameters obtained for BPS\,CS\,22940$-$0009 is given in Table \ref{tab:endtable}. The star's physical properties lie between the helium-rich hot subdwarfs and the extreme helium stars in g-T space (Fig.\,\ref{fig:param_comp}). In terms of helium abundance, it lies close to the He-sdBs and is notably less H-deficient than EHe stars of similar temperature, such as LS IV +6\textdegree2. The star has a luminosity class of V \citep{jeffery20}, compared to true hot subdwarfs which are of classes VI-VII \citep{drilling13}. It therefore can be considered a particularly hot, high-gravity extreme helium star, with the closest counterpart being LS IV +6\textdegree2 \citep{jeffery98}. If BPS\,CS\,22940$-$0009 formed from a double helium white dwarf merger, then it is likely to continue to evolve toward the helium main sequence and become a He-sdB \citep{saio00}.

\begin{table}
    \centering
    \begin{tabular}{c|c|c}
        \hline
        Property & Value & Error \\
        \hline
        $T_{\rm eff}$ (K) & 34\,970 & 370\\
        $\log g$ (cm\,s$^{-2}$) & 4.79 & 0.17 \\
        $n_{\rm He}$ & 0.978 & 0.006 \\
        $p$ (mas) & 0.47 & 0.04 \\
        $d$ (kpc) & 2.10 & $^{+0.19}_{-0.16}$ \\[.1cm]
        $R/{\rm R_\odot}$ & 0.42 & ${\pm0.04}$ \\[.1cm]
        $L/{\rm L_\odot}$ & 230 &  $^{+50}_{-40}$ \\[.1cm]
        $M/{\rm M_\odot}$ (median) & 0.39 & $^{+0.08}_{-0.06}$ \\[.1cm]
        $v_{\rm rad}$ (km\,s$^{-1}$) & 32.7 & 2.0 \\
        $v_{\rm t}$ (km\,s$^{-1}$) & 7.8 & 0.2 \\
        $v\sin{i}$ (km\,s$^{-1}$) & 15.0 & 3.3 \\
        \hline
    \end{tabular}
    \caption{Atmospheric and stellar parameters of BPS\,CS\,22940$-$0009.}
    \label{tab:endtable}
\end{table}

\subsection{Abundances}

\begin{figure*}
    \centering
    \includegraphics[width=11.4cm]{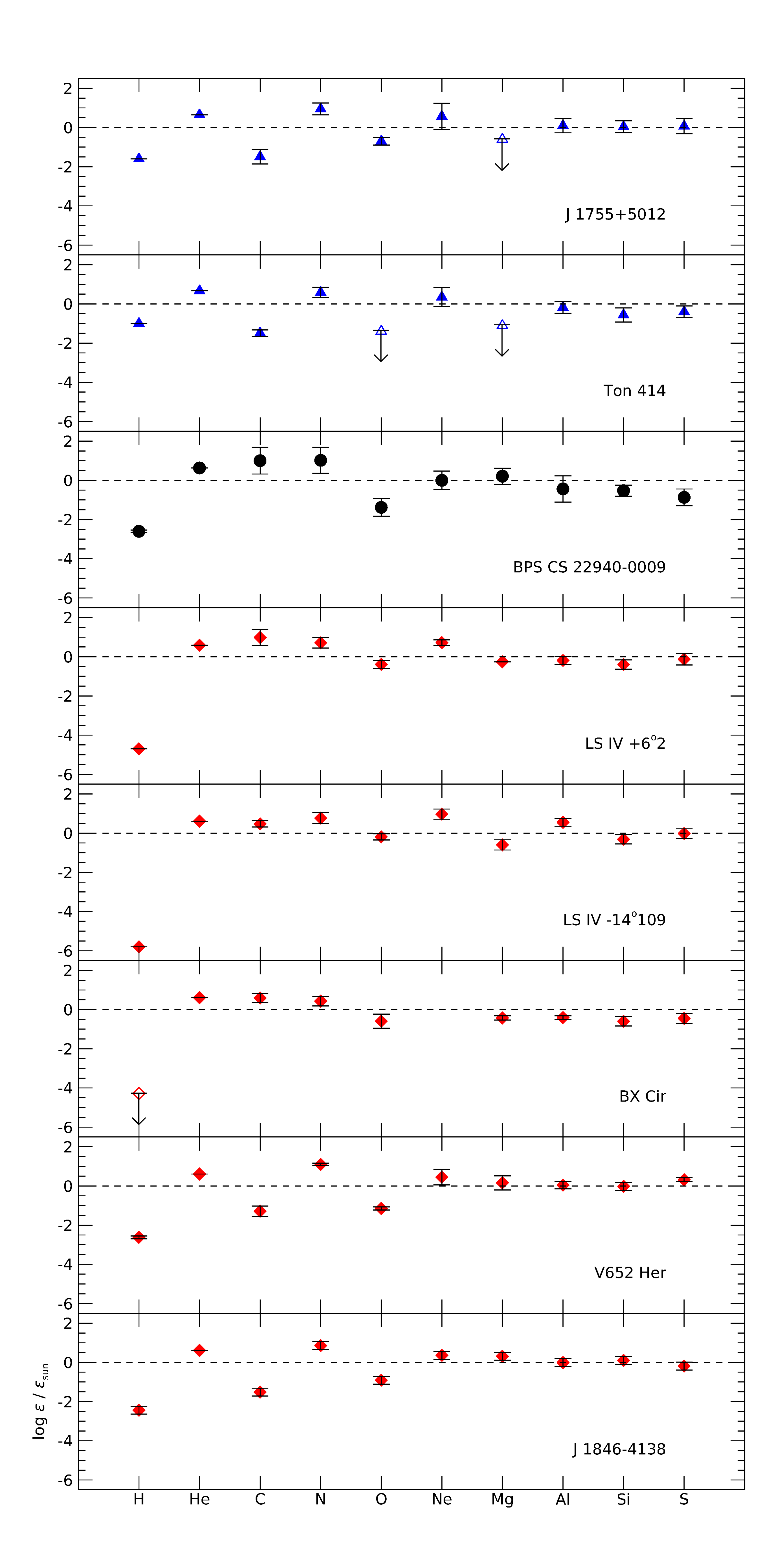}
    \caption{Comparison of the chemical abundances of BPS\,CS\,22940$-$0009 (black circles) against two helium-rich subdwarfs (blue triangles) \citep{naslim20} and five extreme helium stars (red diamonds) \citep{pandey11,jeffery98,pandey06,jeffery99,drilling98,jeffery17b}). Downward arrows indicate upper limits. Solar abundances taken from \citet{asplund09}. Comparison stars were selected from objects with comparable $T_{\rm eff}$ and $\log g$ to BPS\,CS\,22940$-$0009 and available abundances of the same species measurable from the HRS spectrum.} 
    \label{fig:abund_comp}
\end{figure*}

The CNO profile of BPS\,CS\,22940$-$0009 resembles that of extreme helium stars such as BX Cir \citep{jeffery99} and LS IV +6\textdegree2 \citep{jeffery98}. The depletion of H and O coupled with the enhancement of He and N indicate that the stellar material has been CNO-processed. The high C abundance also suggests some triple-$\alpha$ processing. During the merger of two helium white dwarfs, buried carbon-rich material can be dredged to the surface by opacity-driven convection caused by He shell flashes \citep{zhang12}.

The abundances of Al, Si, and S are all significantly subsolar by $\sim0.4-0.9$ dex, while the upper limit for Fe is subsolar by $\sim0.8$ dex. The Ne abundance is approximately solar, which is low compared to other extreme helium stars which also show enhanced carbon (e.g. LS\,IV$+6^{\circ}2$ and LS\,IV$-14^{\circ}109$: Fig.\,\ref{fig:abund_comp}). The ratio of Ne to light elements Al, Si, and S is in line with measurements for C-weak helium stars and subdwarfs (e.g. GALEX J175548.5+501210 (J1755+5012), Ton\,414, V652\,Her and GALEX J184559.8$-$413827 (J1846$-$4138): Fig.\,\ref{fig:abund_comp}). This suggests that the Ne abundance has not been significantly enhanced, for example by $\alpha$-captures on $^{14}$N at the time the excess carbon was produced.

The low metallicity could alternatively be caused by diffusion and stratification of these elements. For stratification to be significant, the star must be long-lived during its current evolution stage with respect to diffusion timescales ($\sim10^5$ yr). This means the star must lie on the helium main sequence or the extended horizontal branch and so must have subdwarf-like surface gravity, which BPS\,CS\,22940$-$0009 does not. Convection currents from He flashes would also disrupt the effects of any diffusion that existed prior to the flash \citep{byrne18}. Therefore if the star is currently evolving towards the helium main sequence, the low metallicity is likely to be intrinsic and not caused by diffusion.

\subsection{Effects of the local thermodynamic equilibrium approximation}
\label{sec:nlte}

\begin{figure}
    \centering
    \includegraphics[width=\columnwidth]{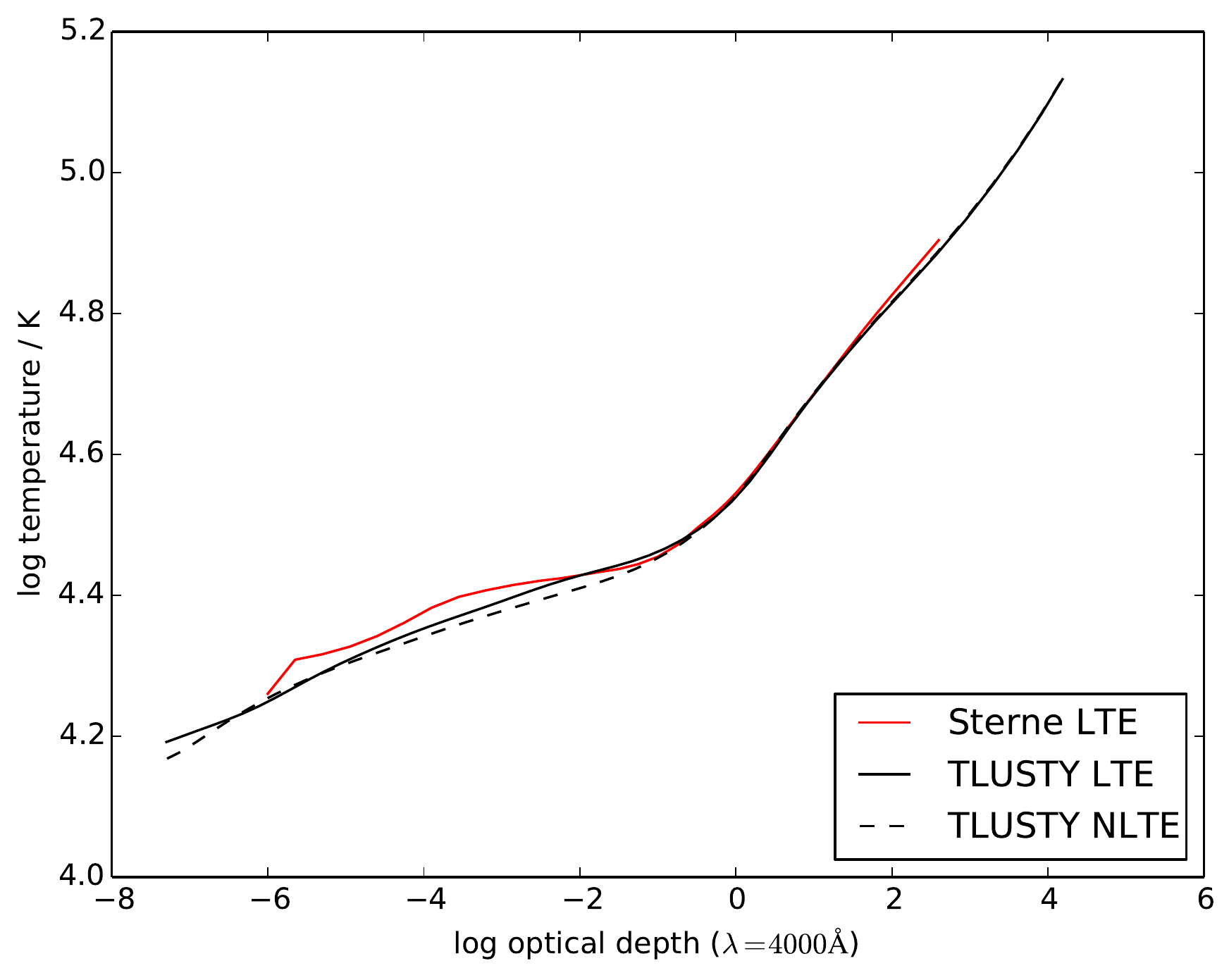}
    \caption{Temperature against monochromatic optical depth for {\sc sterne} and {\sc tlusty} models having $T_{\rm eff}=34\,000$\,K, $\log g/{\rm{cm\,s^{-2}}}=4.75$, $n_{\rm{He}}=0.99$, and $v_{\rm t}=5{\rm km\,s^{-1}}$.}
    \label{fig:tempstruc}.
\end{figure}

For reasons of familiarity with the model atmosphere codes  {\sc sterne} and {\sc spectrum}, our analysis of BPS\,CS\,22940-0009 has assumed the approximation of local thermodynamic equilibrium (LTE) for the determination of electron-level populations used in the equation of state and opacity calculation.
For stars with low surface gravities (or densities) or high effective temperatures, this approximation becomes less secure as the local radiation field increasingly perturbs the population distribution.
The boundary beyond which the LTE approximation is deemed inappropriate has been discussed severally \citep{anderson91,grigsby92,nieva07,pereira11}.
For hot subdwarfs LTE appears satisfactory for $T_{\rm{eff}}< 30\,000$\,K and may be useful up to $T_{\rm{eff}}< 40\,000$\,K (Rauch 2019, private communication).
\citet{jeffery20} found good agreement between effective temperatures and surface gravities using grids of model atmospheres computed with and without the LTE approximation up to   $T_{\rm{eff}}< 40\,000$\,K.
Rather, the use or neglect of the correct line opacity has a far greater influence on the result \citep{anderson91}.
However, this boundary must shift to lower temperatures as surface gravity (or density) is reduced.
In the present case, the surface gravity exceeds that of main-sequence stars, and so should be satisfactory, but is significantly lower than that of most hot “subdwarfs”, and so should be tested.
For the purpose of validating the LTE analysis, we examine how LTE and non-LTE models differ for the properties of this star.

To this end we have computed model atmospheres and emergent spectra with the codes {\sc tlusty}  and {\sc synspec} (version 208) \citep{hubeny17a,hubeny17b,hubeny17c,hubeny21}, including metals up to zinc at the same abundances as the {\sc sterne} models.
We have computed models both with and without LTE  in order to avoid systematic differences  arising from idiosyncrasies peculiar to either  {\sc tlusty} and {\sc synspec} or to  {\sc sterne} and  {\sc spectrum}. The models were computed at 34\,000\,K with a $\log{g}$/cm\,s$^{-2}$ of 4.75, a helium number fraction of 99\% and a $v_{\rm t}$ of 5\,km\,s$^{-1}$.

The distribution of temperature with optical depth is shown in Fig.\,\ref{fig:tempstruc}.
The three models shown are similar above $\log{\tau}>0$, but begin to diverge in the higher layers of the atmosphere. Systematic differences between the {\sc sterne} and {\sc tlusty} LTE models, such as the different treatment of opacities and opacity sampling, contribute to the relatively hotter outer layers of the {\sc sterne} model.

The flux in the NLTE model is $\sim$4\% lower in the region for which broadband photometry data are available (3\,000 to 50\,000\,\AA). This effect would contribute to a $\sim$4\% lower luminosity and a $\sim$2\% lower radius estimation.

In the NLTE model, the neutral He line profiles are both deeper in the core and have less flux in the wings. An NLTE analysis with these models would likely measure a lower surface gravity, but only by $<0.25$ dex. The temperature would likewise have been measured slightly lower, by $<1000$\,K, as the NLTE model has deeper He\,{\sc II} lines and the singly ionised states of carbon and nitrogen are underpopulated compared to the doubly ionised states. This would also affect the abundance measurement of these species, leading to an estimated increase in abundance of $\sim20\%$, given the values of the departure coefficients in the line-forming region.

Whilst models with non-LTE ion populations will certainly improve the fits to most H and He\,{\sc i} line profiles, it is not yet clear that all discrepancies in the fits will be resolved. Further work might include better theoretical profiles in {\sc synspec}, with new line broadening tables and consideration of other potential systematic errors such as departures from plane-parallel geometry.

This brief investigation of LTE vs. NLTE and {\sc tlusty} vs. {\sc sterne} has provided an order of magnitude estimate of the systematic uncertainty which arises from the choice of physics. It has not indicated what the optimum approach might be, though the hybrid LTE model with NLTE formal solution recommended by \citet{nieva07} may prove applicable here also.

\section{Conclusions}

We have presented a detailed analysis of optical spectra of BPS\,CS\,22940$-$0009 to investigate its properties and compare it to similar stars. Precise measurements of the atmospheric parameters and surface abundances have been obtained using grids of line-blanketed, LTE model atmospheres. We found BPS\,CS\,22940$-$0009 to lie on the boundary between the He-sdB and EHe stars in $g$-$T$ space, connecting the two  classes. Due to its luminosity class, the star should not be considered as a true helium-rich subdwarf, but rather a particularly hot, high-gravity extreme helium star. The surface abundances show a composition of CNO-processed material with an intrinsically low metallicity. Strong carbon enhancement suggests that the star formed from a high-mass composite merger of two helium white dwarfs. BPS\,CS\,22940$-$0009 is likely now in a post-merger process of evolving towards the helium main sequence and becoming a true He-sdB star. 

This conclusion would fail were BPS\,CS\,22940$-$0009 to be a member of a close binary. There is no evidence for a cool companion in the SED. We found no evidence of periodic variability in the {\it TESS} photometry for periods $\leq4$\,d. Our attempts to identify variability in the radial velocity (e.g. due to binarity) were inconclusive. 

As a footnote (Section \ref{sec:nlte}), we investigated the impact of the LTE assumption on our analysis, and infer that a modest reduction in $T_{\rm eff}$ and $g$ and a modest increase in elemental abundances would result from a non-LTE analysis. However the differences were comparable with those obtained when comparing LTE models computed with two different model atmosphere codes. Consequently, the LTE assumption has little effect on our overall conclusion.

\section*{Acknowledgements}

The authors thank Andreas Irrgang and Matti Dorsch for the use of and assistance with the implementation of the {\sc isis} SED fitting routines and  Matti Dorsch for advice on the use of {\sc tlusty}. 
They thank the referee for constructive remarks which have led to substantial revision of the paper. 

Some observations reported in this paper were obtained with the Southern African Large Telescope (SALT). This paper includes data collected by the TESS mission. Funding for the TESS mission is provided by the NASA's Science Mission Directorate.
This research has made use of {\sc isis} functions ({\sc isisscripts}) provided by ECAP/Remeis observatory and MIT (http://www.sternwarte.uni-erlangen.de/isis/).
This research has made use of the VizieR catalogue access tool, CDS,
 Strasbourg, France \citep{vizier}.

EJS is supported by the United Kingdom (UK) Science and Technology Facilities Council (STFC) via UK Research and Innovations (UKRI) doctoral training grant ST/R504609/1. 
LJAS and CSJ are supported by the STFC via UKRI research grant ST/V000438/1. 
The Armagh Observatory and Planetarium (AOP) is funded by direct grant from the Northern Ireland Dept for Communities. 
This funding also provides for AOP membership of the United Kingdom SALT consortium (UKSC). 
For the purpose of open access, the authors have applied a Creative Commons Attribution (CC BY) license to any Author Accepted Manuscript version arising.

\section*{Data Availability}

The raw and pipeline reduced SALT observations are available from the SALT Data Archive (https://ssda.saao.ac.za). The model atmospheres and spectra computed for this project are available via the Armagh Observatory and Planetarium web server (https://armagh.space/$\sim$SJeffery/). TESS photometric data are available through the MAST portal (https://mast.stsci.edu).



\bibliographystyle{mnras}
\bibliography{bps_spec} 




\appendix
\renewcommand\thefigure{A.\arabic{figure}} 

\begin{figure*}
    \centering
    \includegraphics[width=\textwidth]{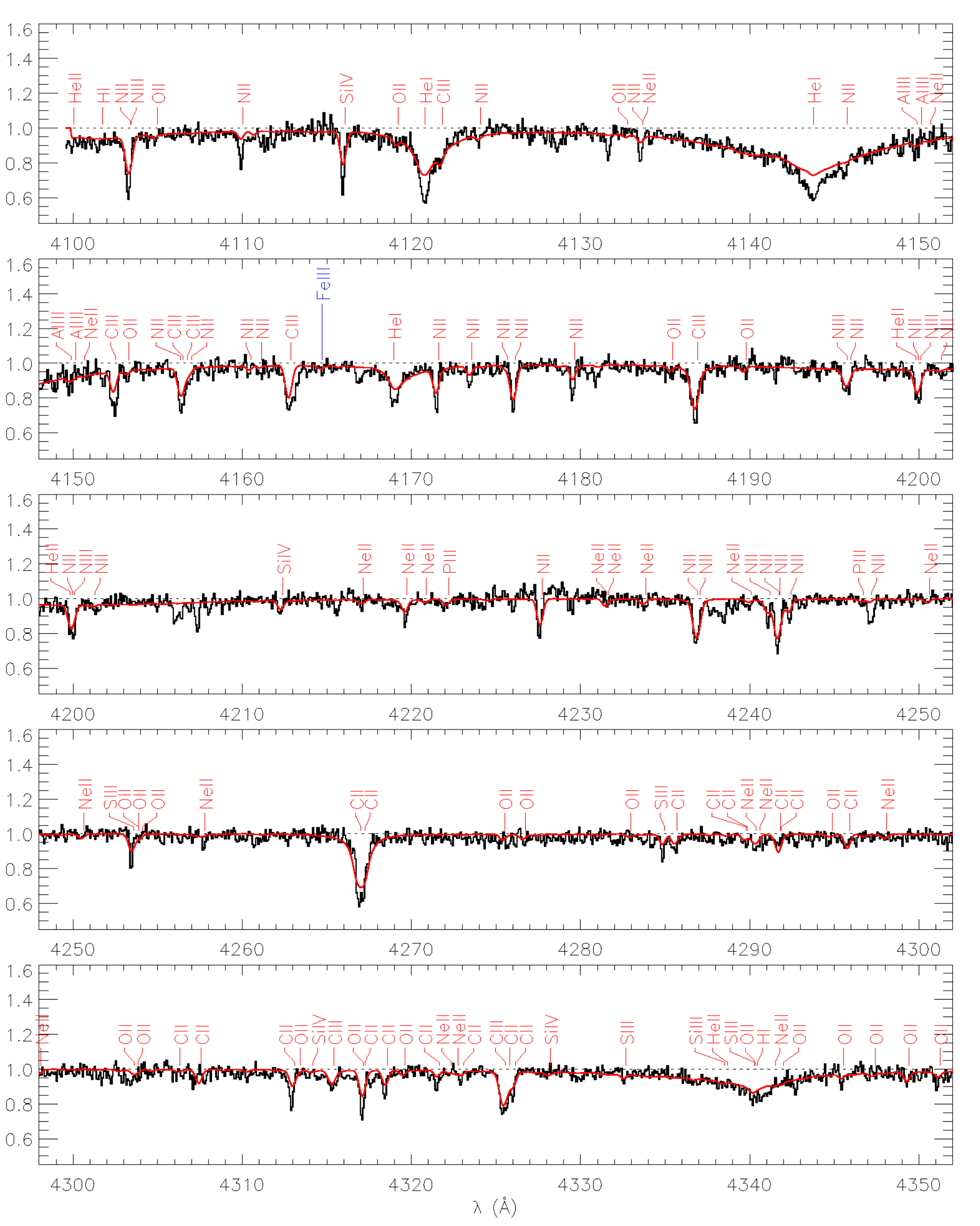}
    \caption{Comparison of the renormalised SALT HRS spectrum of BPS\,CS\,22940$-$0009 (black) and a synthetic spectrum (red) created in {\sc spectrum} using the parameter and abundance results of this analysis.}
    \label{fig:app1a}
\end{figure*}
\begin{figure*}
    \centering
    \includegraphics[width=\textwidth]{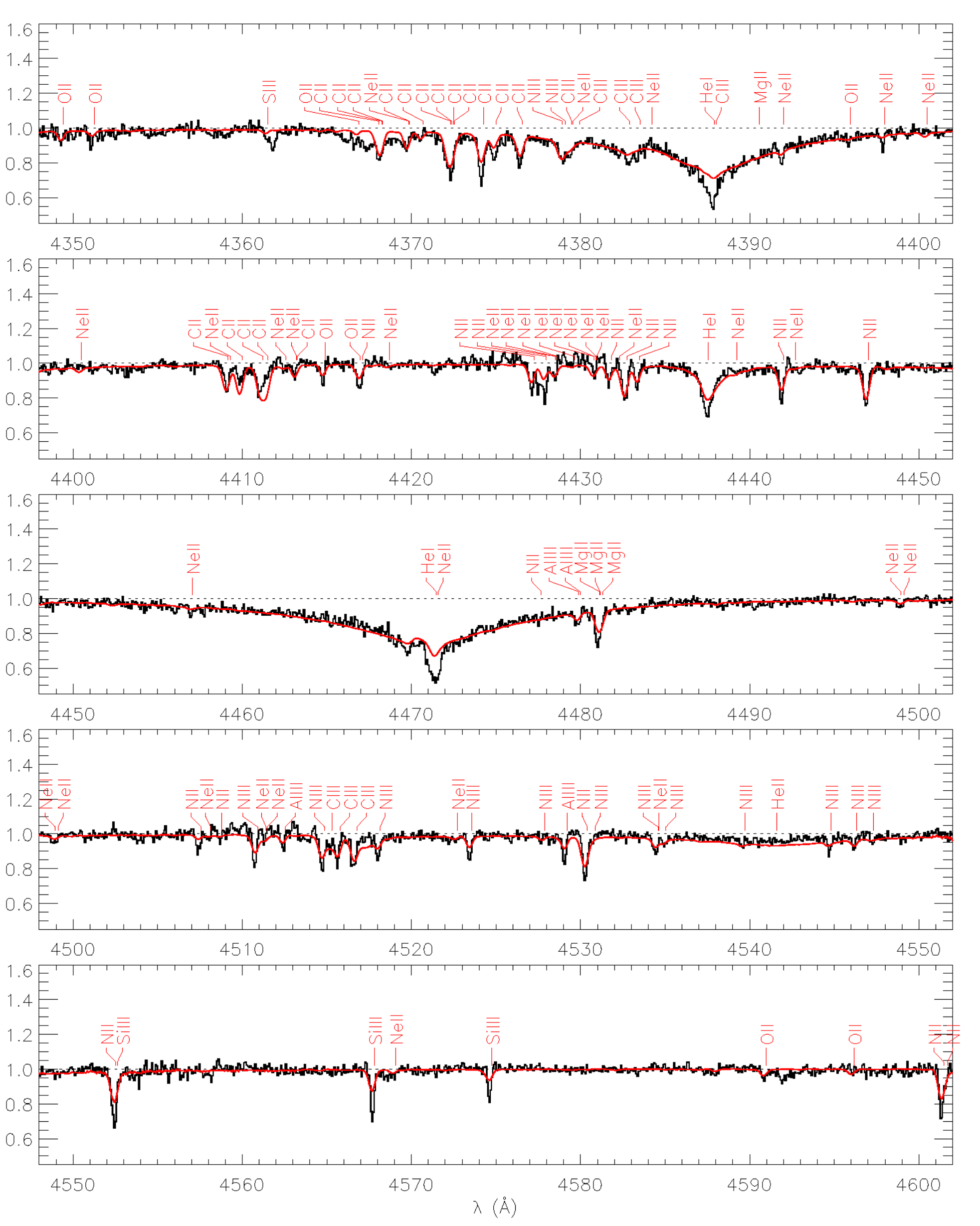}
    \caption{Fig.\ref{fig:app1a} contd.}
    \label{fig:app1b}
\end{figure*}
\begin{figure*}
    \centering
    \includegraphics[width=\textwidth]{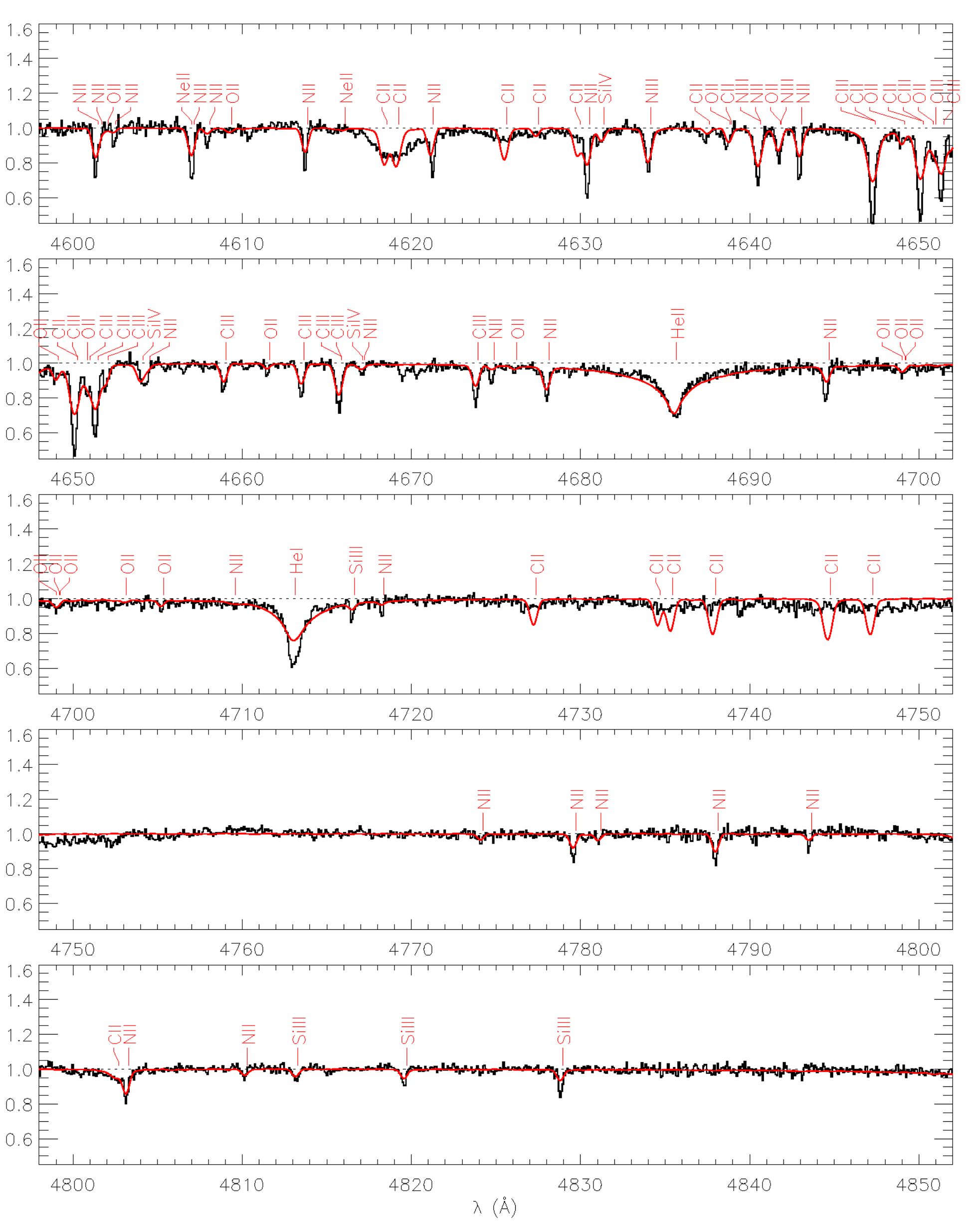}
    \caption{Fig.\ref{fig:app1a} contd.}
    \label{fig:app1c}
\end{figure*}
\begin{figure*}
    \centering
    \includegraphics[width=\textwidth]{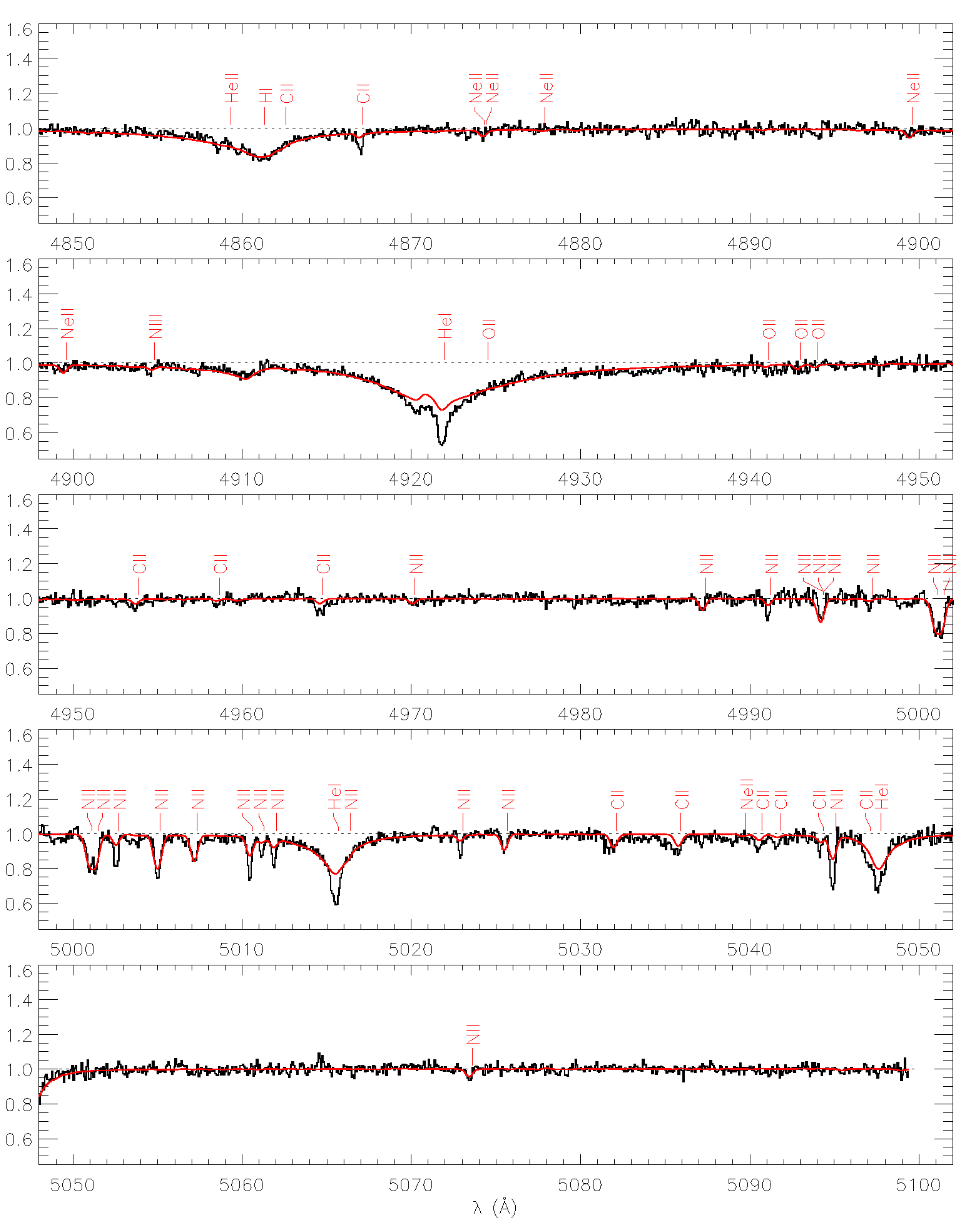}
    \caption{Fig.\ref{fig:app1a} contd.}
    \label{fig:app1d}
\end{figure*}

\section{Renormalised HRS spectrum with formal solution}
\label{app:hrs_spec}

The SALT HRS spectrum of BPS\,CS\,22940$-$0009 and our computed synthetic spectrum are shown in Figs.\ref{fig:app1a}-\ref{fig:app1d}.

\renewcommand\thetable{B.\arabic{table}} 

\begin{table*}
    \centering
    \begin{tabular}{ccccccccc}
                    & Species/             &                &                 & Species/             &                &                 & Species/             &                \\
    $\lambda$ (\AA) & $W_{\lambda}$ (m\AA) & $\log\epsilon$ & $\lambda$ (\AA) & $W_{\lambda}$ (m\AA) & $\log\epsilon$ & $\lambda$ (\AA) & $W_{\lambda}$ (m\AA) & $\log\epsilon$ \\
    \hline
        & H {\sc i}    &                 &         & N {\sc ii} (cont.) &           &         & N {\sc iii}  &                 \\
    4340.46 & 382 $\pm$ 82 & 9.44            & 4227.74 &  52 $\pm$ 30 & 8.54 $\pm$ 6.12 & 4378.93 &  85 $\pm$ 30 & 8.57 $\pm$ 0.33 \\
    4861.32 & 570 $\pm$ 53 & 9.50            & 4236.93 & 103 $\pm$ 37 & 9.04 $\pm$ 0.41 & 4379.11 &  79 $\pm$ 32 & 8.40 $\pm$ 0.38 \\
            & C {\sc ii}   &                 & 4237.05 & 102 $\pm$ 38 & 8.84 $\pm$ 0.46 & 4544.80 &  15 $\pm$  8 & 8.30 $\pm$ 1.41 \\ 
	4267.02 & 287 $\pm$ 74 & 9.50 $\pm$ 0.27 & 4241.18 &  34 $\pm$ 15 & 8.71 $\pm$ 0.30 & 4546.32 &   7 $\pm$  5 & 7.83 $\pm$ 0.25 \\
	4267.27 & 281 $\pm$ 71 & 9.30 $\pm$ 0.26 & 4241.76 & 107 $\pm$ 30 & 8.59 $\pm$ 0.33 & 4547.30 &   7 $\pm$  5 & 8.64 $\pm$ 0.42 \\
	4285.70 &  31 $\pm$ 21 & 9.33 $\pm$ 0.41 & 4241.76 & 104 $\pm$ 28 & 8.56 $\pm$ 0.33 & 4640.64 & 133 $\pm$ 32 & 8.67 $\pm$ 0.61 \\
	4307.58 &  23 $\pm$ 20 & 8.96 $\pm$ 0.50 & 4242.44 &  34 $\pm$ 16 & 8.71 $\pm$ 0.32 & 4640.64 & 131 $\pm$ 31 & 8.65 $\pm$ 0.30 \\
	4313.10 &  74 $\pm$ 29 & 9.87 $\pm$ 0.53 & 4247.22 &  62 $\pm$ 24 & 9.85 $\pm$ 0.36 & 4641.85 &  59 $\pm$ 20 & 8.67 $\pm$ 0.32 \\
	4317.26 &  92 $\pm$ 34 & 9.83 $\pm$ 0.58 & 4417.10 &  48 $\pm$ 19 & 9.02 $\pm$ 0.33 & 4641.85 &  58 $\pm$ 20 & 8.65 $\pm$ 0.34 \\
	4318.60 &  51 $\pm$ 26 & 9.48 $\pm$ 0.49 & 4427.24 &  62 $\pm$ 28 & 8.95 $\pm$ 0.46 &         & O {\sc ii}   &                 \\
	4321.65 &  44 $\pm$ 22 & 9.84 $\pm$ 0.42 & 4427.96 & 129 $\pm$ 48 & 9.99 $\pm$ 0.47 & 4319.63 &  11 $\pm$ 14 & 7.35 $\pm$ 0.61 \\ 
	4323.10 &  16 $\pm$ 13 & 9.39 $\pm$ 0.45 & 4431.82 &  35 $\pm$ 18 & 8.60 $\pm$ 0.37 & 4345.56 &  21 $\pm$ 19 & 7.66 $\pm$ 0.55 \\ 
	4368.27 & 102 $\pm$ 31 & 9.76 $\pm$ 0.40 & 4432.74 &  74 $\pm$ 26 & 8.53 $\pm$ 0.42 & 4349.43 &  10 $\pm$ 14 & 6.87 $\pm$ 0.66 \\ 
	4368.27 & 102 $\pm$ 31 & 9.53 $\pm$ 0.39 & 4433.48 &  34 $\pm$ 18 & 8.46 $\pm$ 0.36 & 4351.26 &  15 $\pm$ 14 & 7.35 $\pm$ 0.49 \\ 
	4369.87 &  54 $\pm$ 20 & 9.36 $\pm$ 0.34 & 4442.02 &  46 $\pm$ 18 & 8.35 $\pm$ 0.31 & 4414.90 &  31 $\pm$ 15 & 7.49 $\pm$ 0.34 \\ 
	4370.69 &   5 $\pm$  6 & 8.46 $\pm$ 0.58 & 4447.03 &  82 $\pm$ 25 & 8.67 $\pm$ 0.48 & 4590.97 &  14 $\pm$  8 & 7.23 $\pm$ 0.31 \\ 
	4372.33 & 129 $\pm$ 32 & 9.67 $\pm$ 0.22 & 4507.56 &  24 $\pm$ 18 & 8.96 $\pm$ 0.43 & 4661.63 &   8 $\pm$  6 & 7.16 $\pm$ 0.34 \\
	4372.33 & 129 $\pm$ 32 & 9.69 $\pm$ 0.29 & 4508.79 &  14 $\pm$ 13 & 9.11 $\pm$ 0.46 & 4699.22 &  19 $\pm$ 11 & 7.87 $\pm$ 0.32 \\
	4372.49 & 130 $\pm$ 32 & 9.62 $\pm$ 0.27 & 4530.40 &  81 $\pm$ 22 & 8.57 $\pm$ 0.79 & 4705.35 &   6 $\pm$  5 & 6.84 $\pm$ 0.39 \\
	4372.49 & 130 $\pm$ 32 & 9.62 $\pm$ 0.29 & 4552.53 & 120 $\pm$ 30 & 9.51 $\pm$ 0.31 &         & Ne {\sc ii}  &                 \\
	4374.27 &  95 $\pm$ 28 & 9.18 $\pm$ 0.25 & 4601.48 &  99 $\pm$ 31 & 9.31 $\pm$ 0.57 & 4219.74 &  41 $\pm$ 22 & 8.00 $\pm$ 0.39 \\
	4375.01 &  58 $\pm$ 21 & 9.40 $\pm$ 0.30 & 4602.53 &  25 $\pm$ 16 & 9.01 $\pm$ 0.38 & 4233.85 &   7 $\pm$ 10 & 7.57 $\pm$ 0.66 \\
	4376.56 &  52 $\pm$ 19 & 8.91 $\pm$ 0.28 & 4607.16 &  91 $\pm$ 28 & 9.27 $\pm$ 0.51 & 4250.65 &   6 $\pm$  8 & 7.51 $\pm$ 0.63 \\
	4409.16 &  42 $\pm$ 18 & 8.94 $\pm$ 0.33 & 4608.09 &  23 $\pm$ 12 & 8.86 $\pm$ 0.30 & 4290.37 &  17 $\pm$ 15 & 7.84 $\pm$ 0.50 \\
	4409.99 &  38 $\pm$ 16 & 8.57 $\pm$ 0.31 & 4613.87 &  61 $\pm$ 20 & 8.80 $\pm$ 0.42 & 4290.60 &  17 $\pm$ 15 & 7.94 $\pm$ 0.48 \\
	4411.20 &  94 $\pm$ 29 & 9.31 $\pm$ 0.36 & 4621.29 &  98 $\pm$ 24 & 9.40 $\pm$ 0.44 & 4369.86 &  34 $\pm$ 24 & 8.54 $\pm$ 0.48 \\
	4411.52 & 108 $\pm$ 22 & 9.31 $\pm$ 0.26 & 4630.54 & 156 $\pm$ 31 & 9.60 $\pm$ 0.26 & 4379.55 &  81 $\pm$ 32 & 8.63 $\pm$ 0.41 \\
	4413.26 &  23 $\pm$ 11 & 9.37 $\pm$ 0.27 & 4643.09 &  86 $\pm$ 24 & 9.09 $\pm$ 0.49 & 4391.99 &  29 $\pm$ 20 & 7.64 $\pm$ 0.43 \\
	4637.63 &  32 $\pm$ 11 & 9.64 $\pm$ 0.23 & 4654.53 &  60 $\pm$ 23 & 9.56 $\pm$ 0.37 & 4397.99 &  20 $\pm$ 13 & 8.21 $\pm$ 0.36 \\
	4638.92 &  51 $\pm$ 19 & 9.66 $\pm$ 0.33 & 4667.21 &  17 $\pm$  8 & 8.78 $\pm$ 0.23 & 4409.30 &  31 $\pm$ 13 & 7.81 $\pm$ 0.28 \\
	4867.07 &  38 $\pm$ 11 & 9.73 $\pm$ 0.22 & 4674.91 &  18 $\pm$ 10 & 8.83 $\pm$ 0.30 & 4413.22 &  27 $\pm$ 13 & 8.01 $\pm$ 0.30 \\
	4953.86 &  34 $\pm$ 17 & 9.75 $\pm$ 0.35 & 4678.14 &  66 $\pm$ 21 & 8.59 $\pm$ 0.33 & 4428.63 &  31 $\pm$ 19 & 8.08 $\pm$ 0.40 \\
	5032.13 &  58 $\pm$ 16 & 9.60 $\pm$ 0.28 & 4694.70 &  73 $\pm$ 30 & 8.97 $\pm$ 0.40 & 4430.79 &  15 $\pm$ 16 & 7.77 $\pm$ 0.58 \\
	5035.94 &  56 $\pm$ 23 & 9.68 $\pm$ 0.40 & 4718.38 &  17 $\pm$  9 & 8.87 $\pm$ 0.29 & 4430.95 &  15 $\pm$ 17 & 8.16 $\pm$ 0.60 \\
	5040.71 &  36 $\pm$ 21 & 9.70 $\pm$ 0.42 & 4774.24 &  11 $\pm$  6 & 8.40 $\pm$ 0.31 & 4511.48 &  18 $\pm$ 19 & 7.88 $\pm$ 0.59 \\
	5044.36 &  23 $\pm$ 20 & 9.31 $\pm$ 0.53 & 4779.72 &  42 $\pm$ 14 & 8.79 $\pm$ 0.29 & 4522.72 &  12 $\pm$  8 & 7.94 $\pm$ 0.36 \\
            & C {\sc iii}  &                 & 4788.13 &  57 $\pm$ 21 & 8.89 $\pm$ 0.41 &         & Mg {\sc ii}  &                 \\
	4152.51 & 110 $\pm$ 50 & 9.92 $\pm$ 0.58 & 4793.65 &  18 $\pm$ 12 & 8.70 $\pm$ 0.37 & 4481.13 &  86 $\pm$ 32 & 7.86 $\pm$ 0.41 \\
	4186.90 & 175 $\pm$ 62 & 9.18 $\pm$ 0.44 & 4803.29 & 131 $\pm$ 27 & 9.80 $\pm$ 0.28 & 4481.13 &  85 $\pm$ 32 & 7.84 $\pm$ 0.41 \\
	4315.44 &  52 $\pm$ 25 & 9.27 $\pm$ 0.39 & 4810.31 &   8 $\pm$  7 & 8.25 $\pm$ 0.46 & 4481.33 &  87 $\pm$ 34 & 7.74 $\pm$ 0.41 \\
	4382.90 &  86 $\pm$ 38 & 9.82 $\pm$ 0.42 & 4970.23 &  15 $\pm$ 10 & 8.71 $\pm$ 0.57 &         & Al {\sc iii} &                 \\
	4515.33 &  29 $\pm$ 17 & 8.98 $\pm$ 0.40 & 4987.38 &  32 $\pm$ 15 & 8.72 $\pm$ 0.34 & 4149.90 &   7 $\pm$ 15 & 5.72 $\pm$ 1.04 \\ 
	4515.78 &  47 $\pm$ 22 & 8.87 $\pm$ 0.33 & 4991.22 &  41 $\pm$ 22 & 9.28 $\pm$ 0.49 & 4150.14 &   5 $\pm$ 14 & 5.72 $\pm$ 1.31 \\
	4516.77 &  65 $\pm$ 27 & 8.93 $\pm$ 0.37 & 4994.35 &  47 $\pm$ 30 & 9.76 $\pm$ 0.53 & 4479.89 &  19 $\pm$  9 & 6.07 $\pm$ 0.27 \\
	4651.01 &  54 $\pm$ 28 & 9.06 $\pm$ 7.20 & 4994.36 &  48 $\pm$ 29 & 9.77 $\pm$ 0.50 & 4479.97 &  17 $\pm$ 11 & 6.02 $\pm$ 0.33 \\
	4651.47 & 180 $\pm$ 50 & 9.36 $\pm$ 0.38 & 4994.37 &  46 $\pm$ 29 & 9.73 $\pm$ 0.51 & 4512.54 &  20 $\pm$ 25 & 6.12 $\pm$ 0.71 \\
	4659.06 &  46 $\pm$ 19 & 9.14 $\pm$ 8.01 & 4997.23 &  16 $\pm$ 12 & 9.12 $\pm$ 0.43 & 4529.20 &  41 $\pm$ 15 & 6.38 $\pm$ 0.33 \\
	4663.64 &  56 $\pm$ 14 & 9.22 $\pm$ 0.26 & 5001.13 & 101 $\pm$ 30 & 9.19 $\pm$ 0.54 &         & Si {\sc iii} &                 \\
	4665.86 & 101 $\pm$ 20 & 9.39 $\pm$ 0.30 & 5001.47 & 112 $\pm$ 35 & 9.20 $\pm$ 0.56 & 4813.30 &  32 $\pm$ 11 & 7.08 $\pm$ 0.22 \\
	4665.86 & 102 $\pm$ 19 & 9.40 $\pm$ 0.28 & 5002.70 &  41 $\pm$ 19 & 8.93 $\pm$ 0.41 & 4819.72 &  32 $\pm$ 11 & 6.96 $\pm$ 0.22 \\
	4673.95 & 101 $\pm$ 20 & 9.54 $\pm$ 0.36 & 5005.15 &  89 $\pm$ 27 & 8.65 $\pm$ 0.54 & 4828.96 &  48 $\pm$ 16 & 7.15 $\pm$ 0.27 \\
            & N {\sc ii}   &                 & 5007.33 &  69 $\pm$ 17 & 8.71 $\pm$ 0.35 &         & Si {\sc iv}  &                 \\
	4133.67 &  37 $\pm$ 22 & 9.04 $\pm$ 9.76 & 5010.62 &  62 $\pm$ 21 & 8.91 $\pm$ 0.45 & 4212.41 &  23 $\pm$ 14 & 6.69 $\pm$ 0.39 \\
	4171.60 &  83 $\pm$ 28 & 8.84 $\pm$ 0.38 & 5011.30 &  20 $\pm$ 12 & 8.75 $\pm$ 0.35 &         & S {\sc iii}  &                 \\
	4173.57 &  36 $\pm$ 20 & 8.81 $\pm$ 0.36 & 5012.03 &  37 $\pm$ 16 & 8.87 $\pm$ 0.36 & 4253.59 &  33 $\pm$ 21 & 6.17 $\pm$ 0.43 \\
	4176.16 &  75 $\pm$ 28 & 8.39 $\pm$ 0.40 & 5023.05 &  36 $\pm$ 16 & 9.17 $\pm$ 0.36 & 4284.98 &  26 $\pm$ 16 & 6.31 $\pm$ 0.38 \\
	4179.67 &  58 $\pm$ 25 & 8.92 $\pm$ 0.36 & 5025.66 &  31 $\pm$ 11 & 8.47 $\pm$ 0.26 & 4332.69 &   7 $\pm$  8 & 5.98 $\pm$ 0.52 \\
	4195.97 &  65 $\pm$ 36 & 9.16 $\pm$ 2.00 & 5045.09 & 100 $\pm$ 38 & 9.44 $\pm$ 0.66 & 4361.53 &  16 $\pm$ 12 & 6.52 $\pm$ 0.38 \\
	4199.98 &  73 $\pm$ 32 & 8.96 $\pm$ 0.47 & 5073.59 &  25 $\pm$ 10 & 8.75 $\pm$ 0.25 &         &              &                 \\
\end{tabular}
    \caption{All spectral line measurements used in the abundance calculations for BPS\,CS\,22940$-$0009. $W_{\lambda}$ is the measured equivalent width for each line, with associated error. $\log\epsilon$ is the abundance result and error calculated from $W_{\lambda}$ by {\sc spectrum}. No errors were reported for the hydrogen lines.}
    \label{tab:linelist}
\end{table*}

\section{Table of line measurements}
\label{app:linelist}

The equivalent widths of all lines used in the measurement of elemental abundances are shown in Table\,\ref{tab:linelist}.

\section{Detection thresholds for absorption lines in stellar spectra}
\label{app:width}

Various approaches to determine the detection limit for weak spectral features may be found \citep[e.g.][]{cayrel88,stetson08}; that adopted here is as follows. 

The spectrum of radiation $f$ emitted by a body as a function of wavelength $\lambda$ is assumed to consist of a continuum $c(\lambda)$ interrupted by absorption lines. Defining $a\equiv f/c$, the equivalent width of an isolated absorption line is
\begin{equation*}
    W_{\lambda}=\int_{\lambda_p}^{\lambda_q}\Big(1-\frac{f(\lambda)}{c(\lambda)}\Big)d\lambda=\int_{\lambda_p}^{\lambda_q}(1-a(\lambda))d\lambda,
\end{equation*}
where the integration limits ${\lambda_p},{\lambda_q}$ bracket the line in question.

The continuum flux $c(\lambda)$ in the vicinity of an absorption line may be approximated by a polynomial fit $c_{0}(\lambda)$ to regions of local spectrum excluding absorption lines.
Supposing that the spectrum is sampled discretely at wavelengths $\lambda_{i}$ (pixels) and at intervals $\delta\lambda$ and rectifying such that $a_{i} = f_{i}/c_{0}$, then 
\begin{equation*}
    W_{\lambda}\approx\sum_{i=p}^{q}(1-a_{i})\delta\lambda.
\end{equation*}

In continuum regions containing N pixels, the mean value $\bar{a}\equiv\langle f_{i}/c_{0}\rangle = 1$ and has a standard deviation
\begin{equation*}
    \sigma_{a}=\sqrt{\frac{1}{N-1}\sum_{i=1}^{N}(a_{i}-\bar{a})^{2}}.
\end{equation*}

If each measurement $a_{i}$ is associated with a measurement error $\sigma_{i}$, 
and assuming errors associated with each pixel are independent and Poissonian, and that $\langle\sigma_{i}\rangle\approx\sigma_{a}$, the associated error in $W_{\lambda}$ is
\begin{equation*}
    \sigma_{W}\approx\sigma_{a}\sqrt\frac{W_{\lambda}}{\delta\lambda}.
\end{equation*}

In general, $\sigma_{i}\approx\sigma_{a}/\sqrt{a_{i}}$. Hence, more precisely, 
\begin{equation*}
    \sigma_{W}\approx\frac{\sigma_{a}}{\sqrt{\langle a_{i}\rangle}}\sqrt{\frac{W_{\lambda}}{\delta\lambda}}.
\end{equation*}

The corollary is that, for weak lines where $\langle a_{i}\rangle \approx 1$, the signal-to-noise ratio $R_{\rm SN}$ for a line of equivalent width $W_{\lambda}$ is given by
\begin{equation*}
    R_{\rm SN}\equiv\frac{W_{\lambda}}{\sigma_{W}}
    \approx \frac{\sqrt{W_{\lambda}\delta\lambda}}{\sigma_{a}}  
\end{equation*}
and hence
\begin{equation*}
    R_{\rm SN}\geq n\;{\rm if}\;W_{\lambda}\geq\frac{n^2\sigma_{a}^2}{\delta\lambda}.
\end{equation*}

The latter provides a threshold for the detection (or non-detection) or a weak line, with $n=1,2$ and $3$ representing confidence levels for detection of 68\%, 90\%, and 99\% respectively.

\renewcommand\thefigure{D.\arabic{figure}} 
\renewcommand\thetable{D.\arabic{table}}


\bsp	
\label{lastpage}
\end{document}